\PassOptionsToPackage{unicode}{hyperref}
\PassOptionsToPackage{hyphens}{url}
\documentclass[
]{article}
\usepackage{amsmath,amssymb}
\usepackage{lmodern}
\usepackage{iftex}
\ifPDFTeX
  \usepackage[T1]{fontenc}
  \usepackage[utf8]{inputenc}
  \usepackage{textcomp} 
\else 
  \usepackage{unicode-math}
  \defaultfontfeatures{Scale=MatchLowercase}
  \defaultfontfeatures[\rmfamily]{Ligatures=TeX,Scale=1}
\fi
\IfFileExists{upquote.sty}{\usepackage{upquote}}{}
\IfFileExists{microtype.sty}{
  \usepackage[]{microtype}
  \UseMicrotypeSet[protrusion]{basicmath} 
}{}
\makeatletter
\@ifundefined{KOMAClassName}{
  \IfFileExists{parskip.sty}{%
    \usepackage{parskip}
  }{
    \setlength{\parindent}{0pt}
    \setlength{\parskip}{6pt plus 2pt minus 1pt}}
}{
  \KOMAoptions{parskip=half}}
\makeatother
\usepackage{xcolor}
\usepackage[margin=1in]{geometry}
\usepackage{longtable,booktabs,array}
\usepackage{calc} 
\usepackage{etoolbox}
\makeatletter
\patchcmd\longtable{\par}{\if@noskipsec\mbox{}\fi\par}{}{}
\makeatother
\IfFileExists{footnotehyper.sty}{\usepackage{footnotehyper}}{\usepackage{footnote}}
\makesavenoteenv{longtable}
\usepackage{graphicx}
\makeatletter
\def\maxwidth{\ifdim\Gin@nat@width>\linewidth\linewidth\else\Gin@nat@width\fi}
\def\maxheight{\ifdim\Gin@nat@height>\textheight\textheight\else\Gin@nat@height\fi}
\makeatother
\setkeys{Gin}{width=\maxwidth,height=\maxheight,keepaspectratio}
\makeatletter
\def\fps@figure{htbp}
\makeatother
\providecommand{\tightlist}{%
  \setlength{\itemsep}{0pt}\setlength{\parskip}{0pt}}
\newlength{\cslhangindent}
\newlength{\csllabelwidth}
\newlength{\cslentryspacingunit} 
\newenvironment{CSLReferences}[2] 
 {
  \setlength{\parindent}{0pt}
  \ifodd #1
  \let\oldpar\par
  \def\par{\hangindent=\cslhangindent\oldpar}
  \fi
  \setlength{\parskip}{#2\cslentryspacingunit}
 }%
 {}
\usepackage{calc}

\usepackage{subfig}
\usepackage{amsmath}
\usepackage{setspace}
\usepackage{bbm}
\usepackage{booktabs}
\usepackage{longtable}
\usepackage{array}
\usepackage{multirow}
\usepackage{wrapfig}
\usepackage{float}
\usepackage{colortbl}
\usepackage{pdflscape}
\usepackage{tabu}
\usepackage{threeparttable}
\usepackage{threeparttablex}
\usepackage[normalem]{ulem}
\usepackage{makecell}
\usepackage{xcolor}
\ifLuaTeX
  \usepackage{selnolig}  
\fi
\IfFileExists{bookmark.sty}{\usepackage{bookmark}}{\usepackage{hyperref}}
\IfFileExists{xurl.sty}{\usepackage{xurl}}{} 
\hypersetup{
  pdftitle={Jointly Estimating Subnational Mortality for Multiple Populations},
  hidelinks,
  pdfcreator={LaTeX via pandoc}}

\title{Jointly Estimating Subnational Mortality for Multiple Populations}
\author{Ameer Dharamshi\(^{1,2}\)\\
Monica Alexander\(^1\)\\
Celeste Winant\(^3\)\\
Magali Barbieri\(^{3,4}\)\\
\strut \\
\(^1\) University of Toronto\\
\(^2\) University of Washington\\
\(^3\) University of California, Berkeley\\
\(^4\) Institut National d'Études Démographiques}
\date{October 04, 2023}

\begin{document}
\maketitle
\begin{abstract}
Understanding patterns in mortality across subpopulations is essential for local health policy decision making. One of the key challenges of subnational mortality rate estimation is the presence of small populations and zero or near zero death counts. When studying differences between subpopulations, this challenge is compounded as the small populations are further divided along socioeconomic or demographic lines. In this paper, we build on principal component-based Bayesian hierarchical approaches for subnational mortality rate estimation to model correlations across subpopulations. The principal components identify structural differences between subpopulations, and coefficient and error models track the correlations between subpopulations over time. We illustrate the use of the model in a simulation study as well as on county-level sex-specific US mortality data. We find that results from the model are reasonable and that it successfully extracts meaningful patterns in US sex-specific mortality. Additionally, we show that ancillary correlation parameters are a useful tool for studying the convergence and divergence of mortality patterns over time.
\end{abstract}

\hypertarget{introduction}{%
\section{Introduction}\label{introduction}}

Reliable mortality estimates are crucial to support health policy design, implementation, and monitoring. Historically, developments in both mortality modelling and understanding mortality differences across populations have focused on national-level estimates for cross-country comparisons. However, as demonstrated by a growing body of evidence (Ezzati 2008; Kindig and Cheng 2013; Murray 2006; Sehgal et al. 2022), patterns in more granular subnational level mortality rates are needed to facilitate local level decisions. This insight has led to a recent line of work focused on studying disparities in mortality outcomes within countries at the subnational level. In the United States, for example, researchers are analyzing differences in life expectancy across state jurisdictions (Woolf and Schoomaker 2019; Montez et al. 2020; Harper, Riddell, and King 2021). At the international level, the United Nations has recently moved to producing estimates of key global health indicators, such as the under-five mortality rate, at subnational levels alongside the longstanding national level estimates (Godwin and Wakefield 2021). In addition, in recent years it has become increasingly clear that disparities in mortality outcomes along demographic dimensions such as sex or gender, race or ethnicity, or socioeconomic status, are growing (Hendi 2015; Crimmins and Zhang 2019; Gutin and Hummer 2021; Masters and Aron 2021). As an example, in the case of child mortality risks, Bhutta (2016) argued that subnational determinants explain a greater portion of mortality than country-level characteristics. In response, estimates of subpopulation mortality risks are needed to identify and understand the mortality patterns of vulnerable groups, and track the effects of policy responses.

In this paper, we focus on sex differences in mortality. Differences in mortality patterns between males and females have long been documented. Women have had longer life expectancy than men in every population and time period where and when female discrimination has not been prevalent. The mortality gap is particularly large at young adult ages, when mortality is mostly determined by external causes (particularly accidents and homicides), but it remains high throughout all working ages. It closes progressively after age 65 to 70 years but never disappears completely. Sex differences in mortality reached a peak in most high-income countries around 1975-1985 but have declined since, in large part because of the convergence in male and female behavior (e.g., smoking) (Gjonça, Tomassini, and Vaupel 1999). In these countries, the difference in life expectancy between men and women ranges from a low of around 3 years in Northern Europe up to nearly 10 years in Eastern Europe. In the United States, the sex difference in life expectancy increased throughout the 20th century to reach a maximum at 7.74 years in 1975 before declining to a low of 4.74 years in 2012. It has increased again since and currently (2021) reaches 5.75 years (Human Mortality Database 2023). The difference varies across states, ranging from 3.7 years (in Utah) to 7.2 years (in the District of Columbia) (United States Mortality Database 2023). Given that the sex gap in life expectancy is highly location-specific, it is of interest to further study the behaviour of sex-specific mortality risks at finer geographic resolutions such as counties. Hence, there is a need for methods that reliably estimate differential mortality risks between males and females (and between other demographic subgroups) in small populations.

While the need for subnational estimates of mortality disaggregated by subpopulation is clear, there are several challenges to obtaining reliable estimates. Although raw death rates can usually be calculated, the natural variations lead to uncertain estimates. It is well understood that subnational mortality estimates are more complex to construct than national level estimates given the large random fluctuations associated with small numbers (Stevenson and Olson 1993). This problem is further compounded when examining mortality across sub-groups within sub-national areas, where inferring underlying mortality risks from raw death rates is challenged by the erratic patterns over age.

Producing reliable estimates of mortality rates and measuring the associated uncertainty requires statistical models that take the natural stochasticity in the data into account. In recent years, an increasing number of studies have demonstrated the usefulness of the Bayesian approach in the construction of such models (see for instance Alexander, Zagheni, and Barbieri (2017); Khana and Warner (2018); Song and Luan (2022)). Building on previous work, in this paper, we propose a general Bayesian model framework to estimate age-specific mortality rates at the subnational level jointly for multiple populations. The model incorporates characteristic shapes of mortality age schedules within a Bayesian hierarchical framework, which allows information on mortality patterns to be shared across populations. We extend previous approaches by accounting for correlation in mortality experiences across subpopulations, rather than assuming subpopulations are independent. The model produces estimates and uncertainty for mortality
rates for each sex. In addition, it estimates higher-level parameters. These parameters summarise trends in different dimensions of mortality over time and show how similar or dissimilar the trends are across groups. We illustrate the model with an application to estimating sex-specific mortality by county in the United States. While we focus on sex-specific mortality, the modelling framework is generalizable to population groups defined by other characteristics (such as race/ethnicity).

The remainder of the paper is structured as follows. We first give a brief overview of recent developments in subnational mortality estimation. We begin our methods discussion by demonstrating that principal components derived from a set of reference mortality curves capture structural differences in subpopulation specific mortality patterns, then follow with a formal statement of the model. Results from the model along with validation exercises using simulated and real sex-specific US county-level mortality data are then provided. Finally, we conclude with a discussion of our findings and identify directions for future research.

\hypertarget{background}{%
\section{Background}\label{background}}

A large body of research exists on small area estimation issues, and recently demographers have increasingly taken advantage of computational advances which make fitting complex statistical models to small-scale mortality data feasible. In particular, there has been a notable increase in the use of Bayesian methods in demographic estimation, particularly subnational estimation. Bayesian methods are particularly suited to demographic contexts as they provide a useful framework to incorporate different data sources in the same model, account for various types of uncertainty, and allow for information exchange across time and space (Bijak and Bryant 2016).

One area of previous work has focused on estimating aggregate indicators at the subnational level, such as child mortality and life expectancy (Mercer et al. 2015; Ševčíková and Raftery 2021). Models on aggregate indicators generally involve temporal and spatial smoothing, allowing for information in mortality trends to be shared across these dimensions. In some cases, models rely on covariates (such as education or income) to stabilize mortality rate estimates from noisy data (Wang et al. 2013; Arias et al. 2018; Murray 2006).

In addition to aggregate mortality indicators, research has focused on producing estimates of age-specific mortality rates, which can form the basis of estimating subnational life tables. Recent advances in this area build on classical demographic approaches of model life tables and relational models, which identify key patterns in mortality over age across a wide range of populations, and allow patterns to be shifted based on a reduced set of parameters (Coale, Demeny, and Vaughan 1983). For example, TOPALS models consist of a standard age schedule and population-specific deviations away from that standard, which are smoothed using linear splines (de Beer 2012). TOPALS-type models have been used to produce subnational estimates of migration and mortality in varying data quality contexts (Gonzaga and Schmertmann 2016; Schmertmann and Gonzaga 2018; Dyrting 2020). TOPALS models do, however, require selecting a single standard mortality curve. This is challenging when studying multiple populations, as each may have a distinct mortality profile.

A related modelling approach derives a set of `principal components' from reference mortality curves which are then used as the basis of a regression framework. This allows a large set of plausible mortality curves to be estimated using a reduced set of parameters. This approach is in the spirit of the Lee-Carter model for mortality forecasting and related work (Lee and Carter 1992). For example, Clark (2019) uses this approach to formulate a new set of model age schedules in data-sparse contexts. In particular, this paper extends an earlier paper by Alexander, Zagheni, and Barbieri (2017) which introduces a Bayesian hierarchical framework that builds on principal components derived from national mortality schedules for use at the subnational level. In more recent work, Alexander and Alkema (2022) have used a principal components approach to model mortality in the broader context of subnational population estimation using a cohort component projection framework.

In general, most existing mortality models inherently assume subpopulations are independent. They model each population separately, then combine or interpret after modelling. However, ideally we would model all subpopulations within one framework as mortality is generally correlated over groups. Existing joint models focus on modelling mortality by sex as it is well understood that changes in male and female mortality are correlated (Noymer and Van 2014). At the national level, sex-specific life expectancy estimates have been produced by the UN using gap-based approaches such as in Raftery, Lalic, and Gerland (2014). Under this approach, female life expectancy is estimated with standard one-sex methods. Estimates of male life expectancy are then constructed by modelling the gap between female and male life expectancy. In the present case, we are interested in estimates of all age-specific mortality rates, not just life expectancy, and a purely gap-based approach to this problem at smaller scales may not be appropriate due to the small counts in both areas and age groups. Additionally, gap-based approaches require the selection of an anchor population such as females in the case of sex-specific modelling. This works in situations where the composition within a population is roughly balanced, but could be problematic when considering variables such as race or ethnicity that vary dramatically across jurisdictions, particularly if the anchor population is near absent in certain regions.

At the subnational level, Rau and Schmertmann (2020) jointly model age- and sex-specific mortality in regions across Germany using a Bayesian TOPALS approach. With regards to sex differences, they obtain a typical pattern of age-specific sex differences across aggregated regions of Germany, and then use this information as a basis for a prior for differences at smaller scales, penalizing large deviations away from the observed differences by sex. In this paper, we take a different approach, and flip the viewpoint from thinking about sex differences, to thinking about covariation in mortality by sex (or any other population subgroups), and explicitly allow for group mortality rates to move together.

\hypertarget{methods}{%
\section{Methods}\label{methods}}

To estimate subnational mortality rates while also extracting patterns across key subpopulations, we propose a Bayesian hierarchical model that builds on a principal component-based approach, and incorporates structures that capture subpopulation interactions. Before defining the model equations, we first discuss why principal components are an attractive modelling strategy in this context.

\hypertarget{motivation}{%
\subsection{Principal components models}\label{motivation}}

The use of principal components is motivated by the fact that age-specific mortality rates tend to display strong regularities across different populations. This means that systematic variation in mortality rates is well captured by a reduced set of parameters which can be modelled such that information is shared across space and time.

To generate principal components, we compute the singular value decomposition of a collection of regional log-mortality curves. The collection of log-mortality curves, \(\mathbf{X}\), is a \(N \times A\) matrix where \(N\) is the number of region-subpopulation-years under consideration and \(A\) is the number of age groups. For US sex-specific mortality, \(N=6066\) (60 years of state-level sex-specific data) and \(A=19\) (ages \textless1, 1-4, 5-9, \ldots, 75-79, 80-84, 85+). Note that different age groups, such as one-year age groups, could also be considered. The singular value decomposition of \(\mathbf{X}\) is then:
\begin{equation}
\label{eq:svd}
\mathbf{X}=\mathbf{U}\boldsymbol{\Sigma}\mathbf{V}'
\end{equation}

where \(\mathbf{U}\) is the \(6066 \times 19\) matrix of left singular values, \(\boldsymbol{\Sigma}\) is a \(19 \times 19\) diagonal matrix of scaling factors, and \(\mathbf{V}\) is the \(19 \times 19\) matrix of right singular values, which we term `principal components'. These extract key patterns in mortality over age. As the \(\mathbf{X}\) matrix contains data from both male and female populations, the patterns captured by the principal components are shared across sexes.

\begin{figure}

{\centering \includegraphics[width=.75\linewidth]{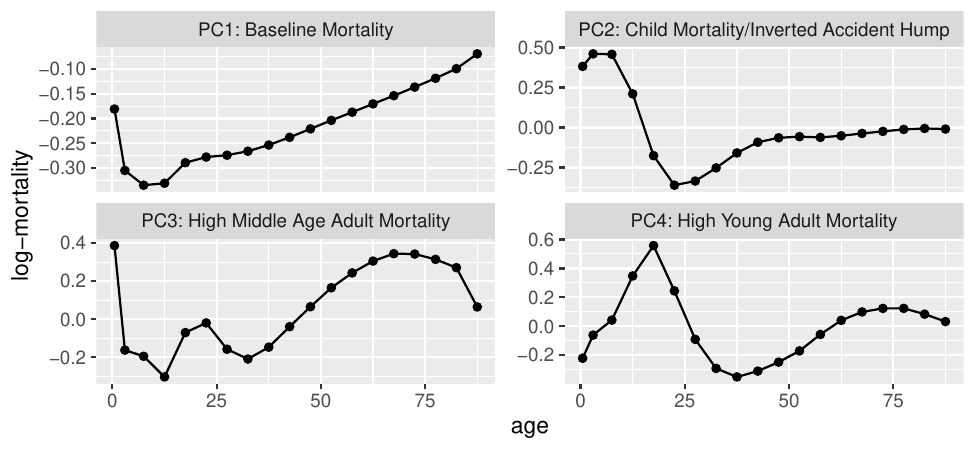} 

}

\caption{First four state-level log-mortality principal components.}\label{fig:pcs}
\end{figure}

The first four principal component curves are plotted in Figure \ref{fig:pcs}. The first principal component explains the most variation across mortality curves with each successive component explaining less variation. The first four principal components explain over 99\% of the variation in the state-level data. Note that the first principal component is inverted for clarity as its coefficients are exclusively negative.

\begin{figure}

{\centering \includegraphics[width=.75\linewidth]{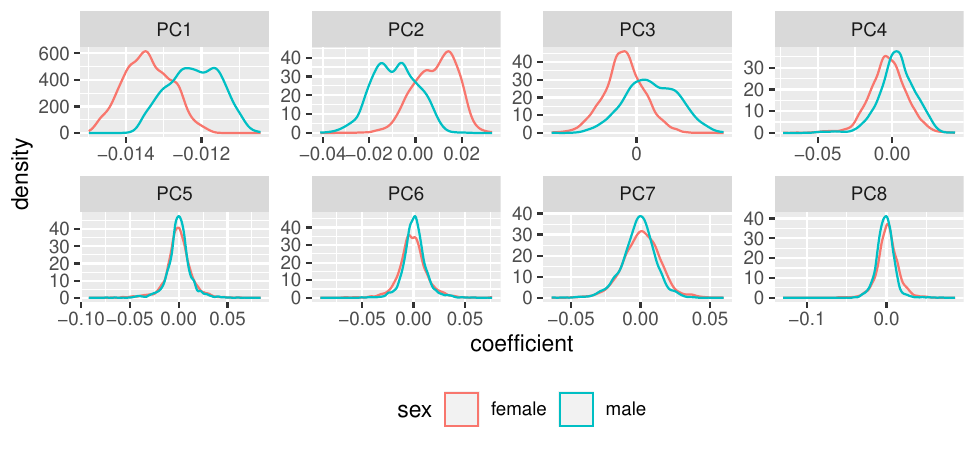} 

}

\caption{Distribution of observed state-level left singular values by sex and principal component.}\label{fig:coefs}
\end{figure}

In practice, choosing the number of components is a balance between incorporating components that only pick up on systematic patterns, while still allowing for enough flexibility in the model. The set of principal components that offer useful information on sex-specific mortality can be identified using the \(\mathbf{U}\) matrix. The left singular values in each row of the \(\mathbf{U}\) matrix represent the contribution of each principal component to the corresponding mortality curve in the \(\mathbf{X}\) matrix. If the \(i\)th principal component contains material subpopulation-specific differences, we would expect the distributions of the values in the \(i\)th column of \(\mathbf{U}\) to differ between subpopulations. We thus separate the rows of \(\mathbf{U}\) by sex and examine the resulting distributions. Figure \ref{fig:coefs} displays the densities of the coefficients by sex for the first eight principal components. We choose eight as the singular values suggest there is limited additional information beyond eight.

The first four principal components demonstrate clear differences in the sex-specific distributions. In particular, the location of the distributions are noticeably different suggesting structural differences between sexes. The distributions of the remaining principal components, including those not presented in Figure \ref{fig:coefs}, are largely similar. Beyond a visual inspection, simple t-tests suggest that aside from the fifth, all of the first eight principal components have location differences. Balancing these findings with the practical consideration that each additional principal component will lead to a significant computational burden during model estimation, we suggest that at least three but ideally four principal components should be used in modelling sex-specific mortality in the US.

Considering Figure \ref{fig:pcs} and Figure \ref{fig:coefs} together offers insight into patterns of interest. The first principal component has the characteristic `J' shape of log mortality. The lower female coefficient values in the PC1 panel of Figure \ref{fig:coefs} indicate that female mortality is generally lower than male mortality. Similarly, the PC3 panel implies that women experience lower mortality through the middle-age adult years. In addition, the distribution of coefficients for the second and fourth principal components indicate higher young adult mortality in the male population than the female population which is consistent with the ``accident hump'' described in the broader mortality literature, as males are more likely to suffer mortality due to risky behaviour (Heligman and Pollard 1980).

There is a substantial body of literature documenting the large and systematic excess mortality of men compared with women (see reviews of this literature in Gjonça, Tomassini, and Vaupel (1999); Barford et al. (2006); Beltrán-Sánchez, Finch, and Crimmins (2015)). As previously mentioned, the male disadvantage in the risk of dying is high at all adult ages up to about 65 to 70 years, depending on the population, and it is particularly acute in young adults. The finding that the principal components are recovering these well understood sex-specific differences motivates their use as the basis for a subpopulation mortality model. Using principal components as a basis for a statistical model can be robust in small population settings as the components allow plausible mortality rates to be estimated even in the presence of high variability.

\hypertarget{model}{%
\subsection{Model summary}\label{model}}

We begin by defining \(y_{a, s, c, t}\) as the observed number of deaths in age group \(a\), subpopulation \(s\), county \(c\), and year \(t\). Then, we assume that
\begin{equation}
\label{eq:pois}
y_{a, s, c, t}|\lambda_{a, s, c, t} \sim \text{Poisson}\left(P_{a, s, c, t} \cdot \lambda_{a, s, c, t} \right),
\end{equation}
where \(P_{a, s, c, t}\) is the population corresponding to age group \(a\), subpopulation \(s\), county \(c\), and year \(t\), and \(\lambda_{a,s,c,t}\) is the mortality rate to be estimated for age group \(a\), subpopulation \(s\), county \(c\), and year \(t\).

The mortality rates \(\lambda_{a,s,c,t}\) are estimated on the log scale as follows:
\begin{equation}
\label{eq:logmort}
\log\left(\lambda_{a,s,c,t}\right) = \sum_{i=1}^P \left(\beta_{i,s,c,t} \cdot Y_{i,a} \right) + \gamma_{a,s,c,t},
\end{equation}

where \(Y_{i}\) is the \(i\)th principal component, \(P\) is the number of principal components, \(\beta\) are the estimated coefficients, and \(\gamma\) is an overdispersion term. The number of principal components, \(P\) can be selected based on their contributions and differences between groups. In the case of sex-specific mortality in US counties, we use the \(P=4\) principal components plotted in Figure \ref{fig:pcs}.

Equation \eqref{eq:logmort} can be intuitively understood as constructing a log-mortality curve for each age-sex-county-year as a linear combination of the four principal components plus some additional age-specific variation. Different linear combinations of the principal components (that is, different estimated values of the \(\beta\) coefficients) lead to different plausible log-mortality curves. The overdispersion term accounts for the possibility that deaths will be overdispersed relative to the patterns captured by the principal components.

\hypertarget{core-model}{%
\subsubsection{Core model}\label{core-model}}

For many counties, population and death counts are small and highly variable, which would lead to uncertain estimates of \(\beta\) if each county and year were estimated independently. As such, we propose a hierarchical model for \(\beta\) based on the nesting structure of counties within states. Specifically, we will assume that counties within each state are more likely to share similar mortality patterns than counties in different states, and allow high-data counties to share information with low-data counties in the same state.
A similar relationship holds across subgroups. Within a county, while subgroups of the population are expected to experience specific drivers of mortality due to their unique experience, they are also expected to jointly experience more general drivers, whether those are due to policy, the environment, or other factors. For instance, regulations regarding maternity leaves have an impact on women's health, but natural catastrophes like heat waves or hurricanes will affect everyone in the same area (albeit to various extents).
This structure suggests that \(\beta\) coefficients, and thus by extension log-mortality curves, for all subgroups within a county should be modelled jointly, thereby allowing for distinct mortality estimates in each group while also exploiting the dependence between groups.

To capture geographic and subgroup dependence, we propose to model the vector of \(\beta\) coefficients for all \(S\) groups within each county jointly as multivariate normal with a common state-level mean vector and covariance matrix:
\begin{align}
\begin{pmatrix} \beta_{i,1,c,t} \\ \dots \\ \beta_{i,S,c,t} \end{pmatrix} &= \begin{pmatrix} \mu_{\beta_{i,1,t}} \\ \dots \\ \mu_{\beta_{i,S,t}} \end{pmatrix} + \begin{pmatrix} \omega_{i,1,c,t} \\ \dots \\ \omega_{i,S,c,t} \end{pmatrix}, \quad i=1,\dots,P \label{eq:beta} \\ 
\left.\begin{pmatrix} \omega_{i,1,c,t} \\ \dots \\ \omega_{i,S,c,t} \end{pmatrix}\right|\sigma_{\beta_{i,t}},L_{i,t}^{(\beta)} &\sim \mathcal{N}\left(\mathbf{0}_S, \sigma_{\beta_{i,t}}\mathbf{1}_S L^{(\beta)}_{i,t} L_{i,t}^{(\beta)\top}\mathbf{1}_S\sigma_{\beta_{i,t}}\right) \\
L^{(\beta)}_{i,t}L_{i,t}^{(\beta)\top} &\sim \text{LKJ}(1) \\
\sigma_{\beta_{i,t}} &\sim \mathcal{N}\left(0,1\right)
\end{align}

where \(\mu_{\beta_{i,s,t}}\) is the state-level coefficient for the \(i\)th principal component, subpopulation \(s\), and year \(t\), and \(\omega_{i,s,c,t}\) is the county deviation for the \(i\)th principal component, subpopulation \(s\), county \(c\), and year \(t\).

We discuss the two components of \(\beta\) in turn, starting with the state-level mean vector, \(\mu_\beta\). Information-rich observations from county-subgroups with large populations are the primary contributors to estimates of \(\mu_\beta\) for the corresponding subpopulation. Estimates of \(\beta\) for small county-subgroups with less informative observed death counts are then partially informed by the high-data counties through the shared state-level means. This pooling effect stabilizes estimates of \(\beta\) for small populations by pulling the estimates towards the state mean. It is important to note that this effect occurs at the county-subgroup level and not uniformly within a county: a large population county with an uneven subgroup composition can experience stronger pooling effects in its smaller subgroups. This is not particularly important for the application to sex-specific mortality as the populations are roughly balanced, but may be critical in other applications involving subgroups defined by other demographic variables.

The \(\omega\) vector models the specific deviations in the \(\beta\) vector for each county from the state-level mean vector. By jointly modelling the deviations for all subpopulations as in \eqref{eq:beta}, the model captures patterns in the dispersion of \(\beta\) by principal component in \(L_{i,t}^{(\beta)}L_{i,t}^{(\beta)\top}\). This enables information sharing across subpopulations. In counties where there is an imbalance in the number of non-zero death observations across subpopulations, jointly generating \(\beta\) coefficients allows higher data groups to support lower data groups. An example of such a county is given in Figure \ref{fig:miss} in Appendix \ref{app-res}.

We assign the uninformative LKJ(1) prior to the subpopulation correlation Cholesky factors to allow the observed data to fully determine the correlation structure (Lewandowski, Kurowicka, and Joe 2009).

By modelling correlations in principal component coefficients between sexes, it is possible for subgroups to experience joint variation in certain parts of their mortality curves but not in others. For example, the coefficients corresponding to the first principal component (ie. baseline mortality) could experience strong correlation, leading to a common trend in baseline mortality levels, but the coefficients on the second principal component (ie. the accident hump) may not, leading to differences by sex in young adult mortality.

\begin{figure}

{\centering \includegraphics[width=.75\linewidth]{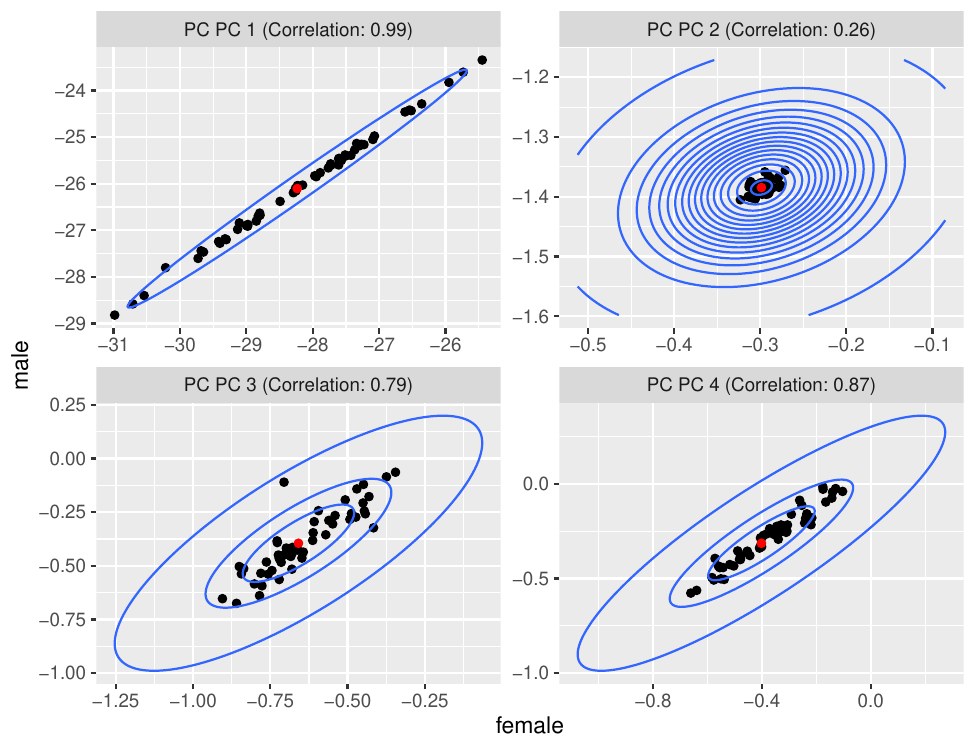} 

}

\caption{Median posterior county-level principal component coefficients ($\beta$) and the corresponding state-level mean ($\mu_\beta$) for California in 2017. The blue contour plots indicate the recovered correlation patters between male and female population.}\label{fig:corrcontours}
\end{figure}

An illustration of the correlated principal component coefficients is provided in Figure \ref{fig:corrcontours}. Each plot captures the patterns of one of the four principal component coefficients for California in 2017. The red point represents the state-level value, the black points represent each county, and the blue contours are the distribution of \(\omega\) centered at the state means. High correlation in the first principal component is expected since when the general conditions in a region improve, both sexes are positively impacted, leading to highly correlated changes in baseline mortality. Similarly, the low correlation in the second principal component is consistent with males experiencing a relatively elevated level of mortality in young adult ages (i.e.~the accident hump) whereas females tend not to. High correlations are recovered for the third and fourth principal components, suggesting that changes in young adult and middle-age adult mortality experiences are similar across sexes.

The \(\beta\) specification is intended to be flexible. If there is only a small number of years of data available or there is prior information suggesting that correlation structures are time-invariant, one could reasonably omit the time index or share correlations across a short time interval. Similarly, if certain correlations are known in advance, the correlation matrices themselves could be constrained accordingly.

\hypertarget{temporal-smoothing}{%
\subsubsection{Temporal smoothing}\label{temporal-smoothing}}

State-level means are smoothed over time by penalizing the second-order differences to produce gradual changes. Smoothing occurs at the state-level as opposed to the county-level to allow the needed flexibility for counties to experience irregular mortality patterns driven by localized events such as a natural disaster, or a pandemic.
\begin{align}
\mu_{\beta_{i,s,t}}|\mu_{\beta_{i,s,t-1}},\mu_{\beta_{i,s,t-2}},\sigma_{\mu_{\beta_i}} &\sim \mathcal{N}\left(2\cdot\mu_{\beta_{i,s,t-1}} - \mu_{\beta_{i,s,t-2}}, \sigma_{\mu_{\beta_{i}}}\right) \\
\sigma_{\mu_{\beta_{i}}} &\sim \mathcal{LN}\left(-1.5, 0.5\right).
\end{align}

\hypertarget{overdispersion-term}{%
\subsubsection{Overdispersion term}\label{overdispersion-term}}

Finally, the \(\gamma_{a,s,c,t}\) term that allows for overdispersion of the log-mortality rate is modelled similarly to \(\omega_{i,s,c,t}\) in a \(S\)-dimensional multivariate normal setup:
\begin{align}
\left.\begin{pmatrix} \gamma_{a,1,c,t} \\ \dots \\ \gamma_{a,S,c,t} \end{pmatrix}\right|\sigma_a,L_{a,t}^{(\gamma)} &\sim \mathcal{N}\left(\mathbf{0}_S, \sigma_{a}\mathbf{1}_S L^{(\gamma)}_{a,t} L_{a,t}^{(\gamma)\top}\mathbf{1}_S \sigma_a\right) \\
\sigma_a &\sim \mathcal{N}\left(0, 0.25\right) \\
L^{(\gamma)}_{a,t}L_{a,t}^{(\gamma)\top} &\sim \text{LKJ}(1).
\end{align}

Relationships between subpopulation \(\gamma\)'s are captured using the age-year correlation matrix \(L^{(\gamma)}_{a,t}L_{a,t}^{(\gamma)\top}\). An intuitive way of understanding this component of the model is that the principal components produce the expected mortality derived from aggregate and local patterns while \(\gamma\) captures additional age-specific deviations. These deviations from the expectation are often correlated, though this correlation may differ across ages and over time. Given that the log-mortality values for different age groups occur in substantially different parts of the log curve, we estimate a separate scaling factor \(\sigma\) for each age group.

\hypertarget{computation}{%
\subsection{Computation}\label{computation}}

The model described here was fit using a Bayesian framework. Posterior samples are drawn using the No-U-turn sampling (NUTS) Hamiltonian Monte Carlo algorithm (Hoffman and Gelman 2014; Neal 2011) implemented in the Stan R package (Stan Development Team 2021). We execute 4 chains with 500 iterations of burn-in and 2 500 iterations of samples. Convergence was diagnosed using trace plots, effective sample sizes, and the Gelman and Rubin diagnostic (Gelman and Rubin 1992).

\hypertarget{simulation-study}{%
\subsection{Simulation study}\label{simulation-study}}

We conducted a simulation study to test the ability of the model to simultaneously estimate mortality rates and extract patterns in subgroup mortality. We generated 10 years of population data for 25 simulated counties of various sizes composed of 5 population subgroups ranging in size from 10\% to 50\% of the total population. We then generated deaths for all subgroups using log-mortality rates constructed as the linear combination of a pair of standard curves with coefficients subject to a variety of correlation patterns. The goal of the simulation study was to recover the true correlations and age-specific mortality rates when fitting the proposed model to the simulated data. Specific details on the data generating process are provided in Appendix \ref{app-sim}.

\begin{figure}

{\centering \subfloat[True correlation matrices\label{fig:simcor-1}]{\includegraphics[width=.45\linewidth]{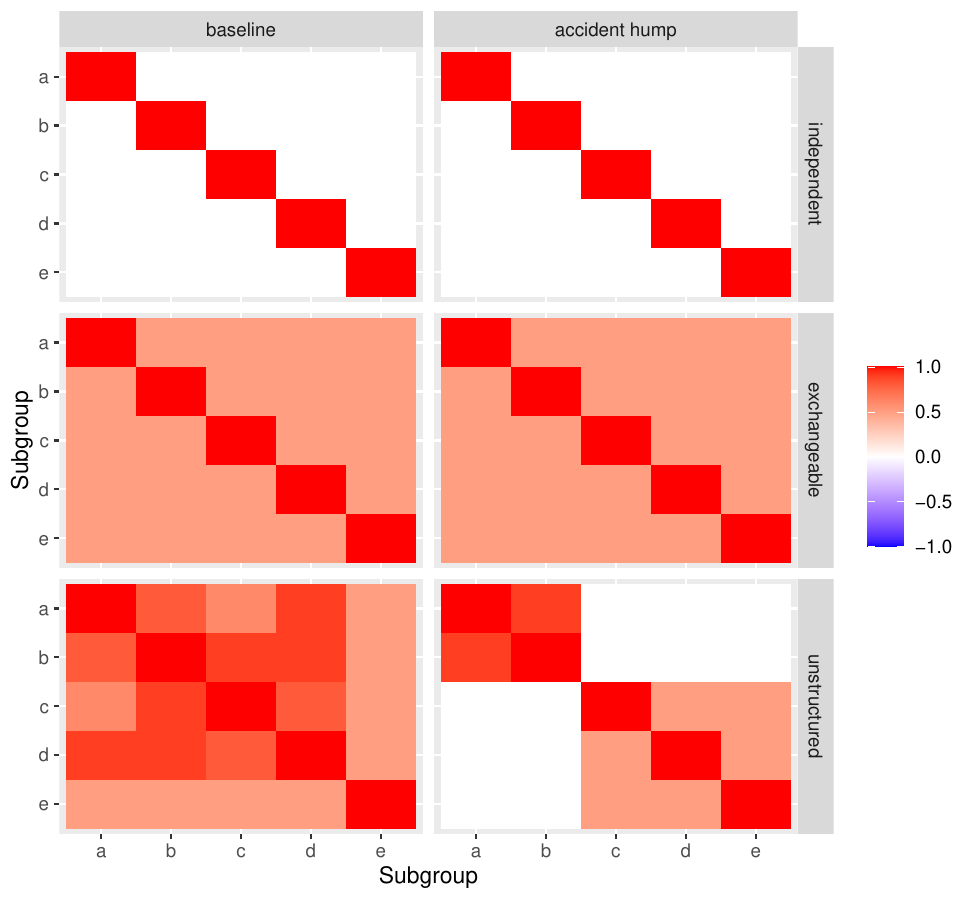} }\subfloat[Estimated correlation matrices\label{fig:simcor-2}]{\includegraphics[width=.45\linewidth]{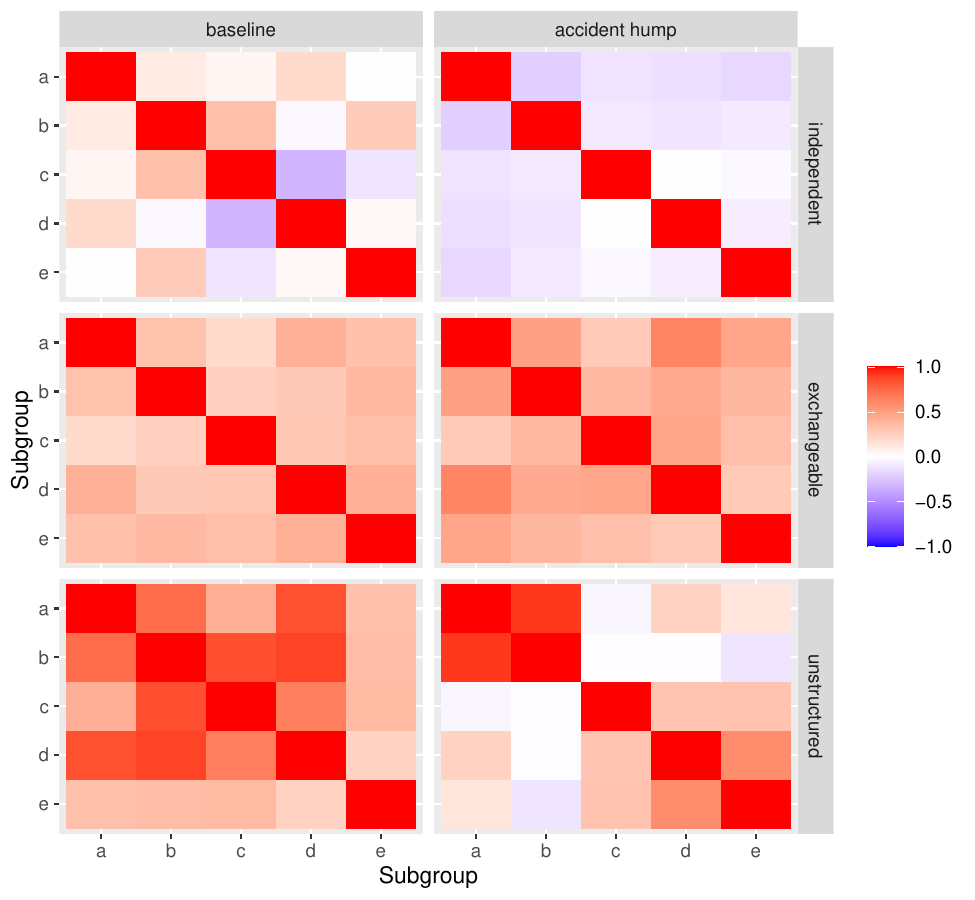} }

}

\caption{Simulation study true and estimated posterior median correlation matrices.}\label{fig:simcor}
\end{figure}

The model is run on the simulated data using the standard curves as the principal components. To validate the results, we compare the estimated correlation matrices and log-mortality rates to the corresponding true values. Figure \ref{fig:simcor}b plots estimated correlation matrices extracted from the model. Comparing corresponding facets against Figure \ref{fig:simcor}a, it appears that the model has successfully identified and extracted the patterns observed in the data.

Table \ref{tab:simtab} presents the coverage values for correlation and log-mortality rate parameters at the 80\%, 90\%, and 95\% nominal levels. Coverage is defined for correlation and log-mortality rates as \(\frac{1}{n}\sum_{i=1}^n\mathbf{1}_{l_i\le \theta_i \le u_i}\) where \(n\) is the number of parameters, \(\theta_i\) refers to an individual parameter, and \(l_i\) and \(u_i\) are respectively the lower and upper bounds of the posterior credible intervals. For the correlation matrices, we compute entrywise coverage values for the off-diagonal entries of the lower triangles of the correlation matrices.

\begin{table}[H]

\caption{\label{tab:simtab}Simulation study coverage values for correlation and log-mortality rates.}
\centering
\begin{tabular}[t]{lrrr}
\toprule
 & Coverage (80\%) & Coverage (90\%) & Coverage (95\%)\\
\midrule
Correlations & 0.78 & 0.90 & 0.94\\
Log-mortality rates & 0.83 & 0.92 & 0.96\\
\bottomrule
\end{tabular}
\end{table}

For both sets of parameters, the model's coverage values are in line with the nominal values at all levels, suggesting that the model is well calibrated.

\hypertarget{application-to-age--and-sex-specific-mortality-rates-at-the-county-level-in-the-united-states}{%
\section{Application to age- and sex-specific mortality rates at the county level in the United States}\label{application-to-age--and-sex-specific-mortality-rates-at-the-county-level-in-the-united-states}}

We applied the model to estimate sex- and age-specific mortality at the county level in the US over the period 1982-2019. In the US, relatively high quality data are collected through vital registration systems. The National Center for Health Statistics (NCHS) publishes detailed mortality records for each year included in our series from which we can compile the mortality tabulations. For purposes of the present analysis, we obtained access to the necessary protected data with individual death records from 1989 to 2019 from the NCHS under a Data User Agreement as the publicly available data do not include geographic information. The United States Census Bureau publishes mid-year population estimates by county of residence, year, sex, age. These data are available to the public through the Census Bureau. To comply with our data agreement, counties are anonymized in all figures.

The model is applied to each state individually using the first four principal components plotted in Figure \ref{fig:pcs}. Note that estimates based on this model are published as part of the United States Mortality DataBase (USMDB) (\url{https://usa.mortality.org/}).

\begin{figure}[htb!]

{\centering \includegraphics[width=0.95\linewidth]{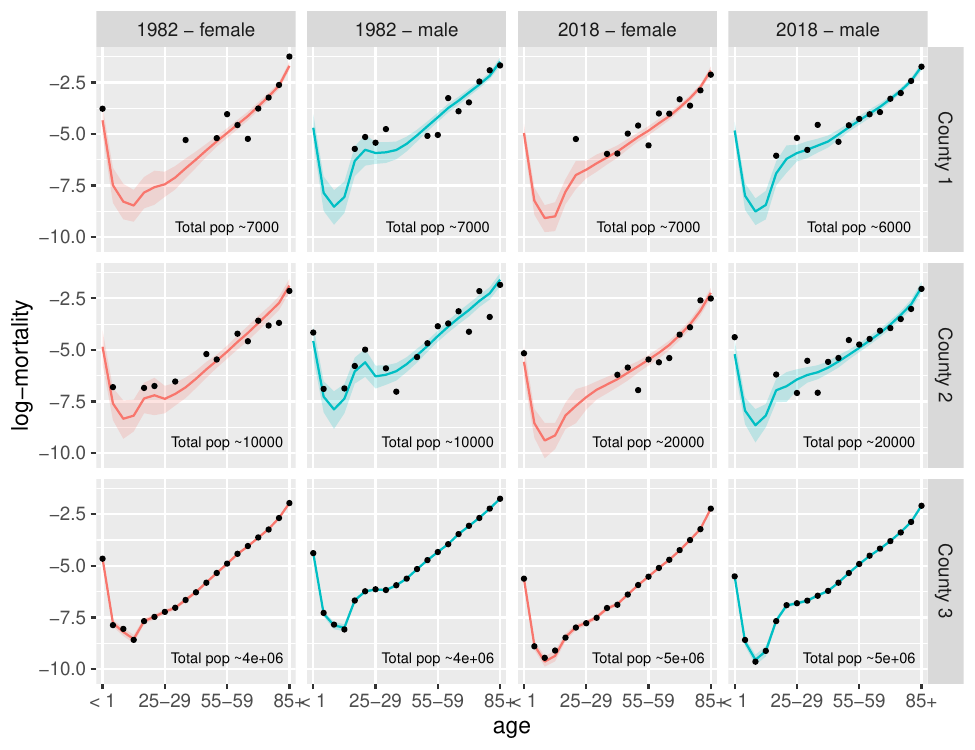} 

}

\caption{Examples of US county-level sex-specific log-mortality plots by county-year-sex for three counties of varying sizes. Black dots represent observed values and coloured curves and regions indicate posterior medians and 95\% credible intervals respectively.}\label{fig:lmx}
\end{figure}

As an illustration of the model outputs, in Figure \ref{fig:lmx} we plot the estimated sex-specific log-mortality curves in 1982 and 2019 for a subset of US counties along with associated uncertainty intervals. The observed log-mortality rates for age groups with non-zero deaths are presented as points, and the female and male county-level estimates and 95\% credible intervals are presented in red and blue respectively. The selected counties represent settings with small, medium, and large populations (of approximately 13,000, 40,000, and 10,000,000 people, respectively). As expected, smaller counties such as County 1 have greater uncertainty as compared to larger ones such as County 3 due to the higher levels of noise caused by low or zero death counts. For County 3, modelled estimates follow the data exactly.

\hypertarget{state-level-patterns-in-sex-specific-mortality}{%
\subsection{State-level patterns in sex-specific mortality}\label{state-level-patterns-in-sex-specific-mortality}}

In addition to producing mortality rates for subnational areas, the proposed model specification contains a number of parameters that are useful in observing broad patterns in mortality at the state-level. Specifically, changes in \(\mu_\beta\) describe broad, state-level trends in mortality patterns over time, and the correlation matrices offer insight into county-level trends.

\begin{figure}

{\centering \includegraphics{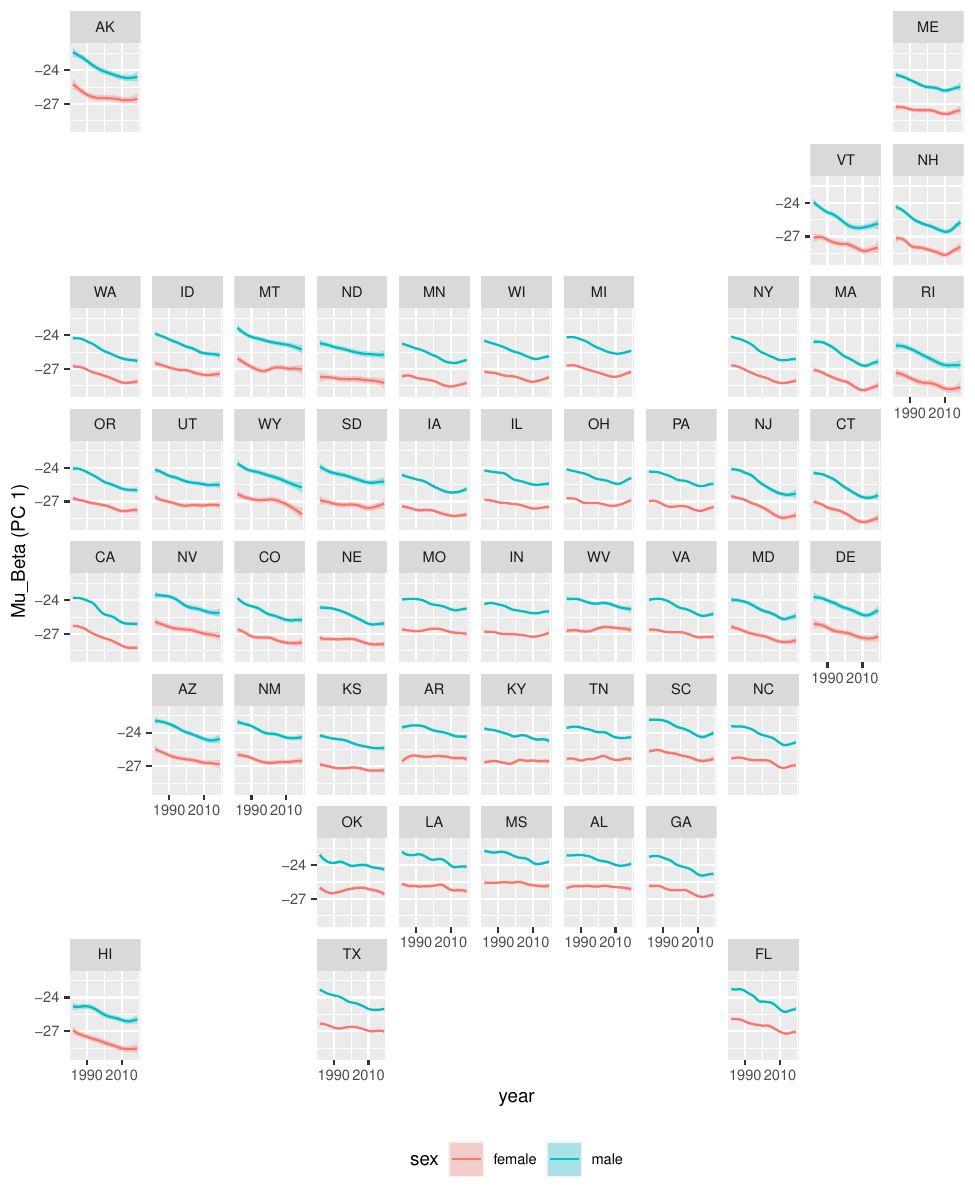} 

}

\caption{Posterior medians and 95\% credible intervals for the state-level coefficients for the first principal component ($\mu_\beta$).}\label{fig:mubetamapPC1}
\end{figure}

\begin{figure}

{\centering \includegraphics{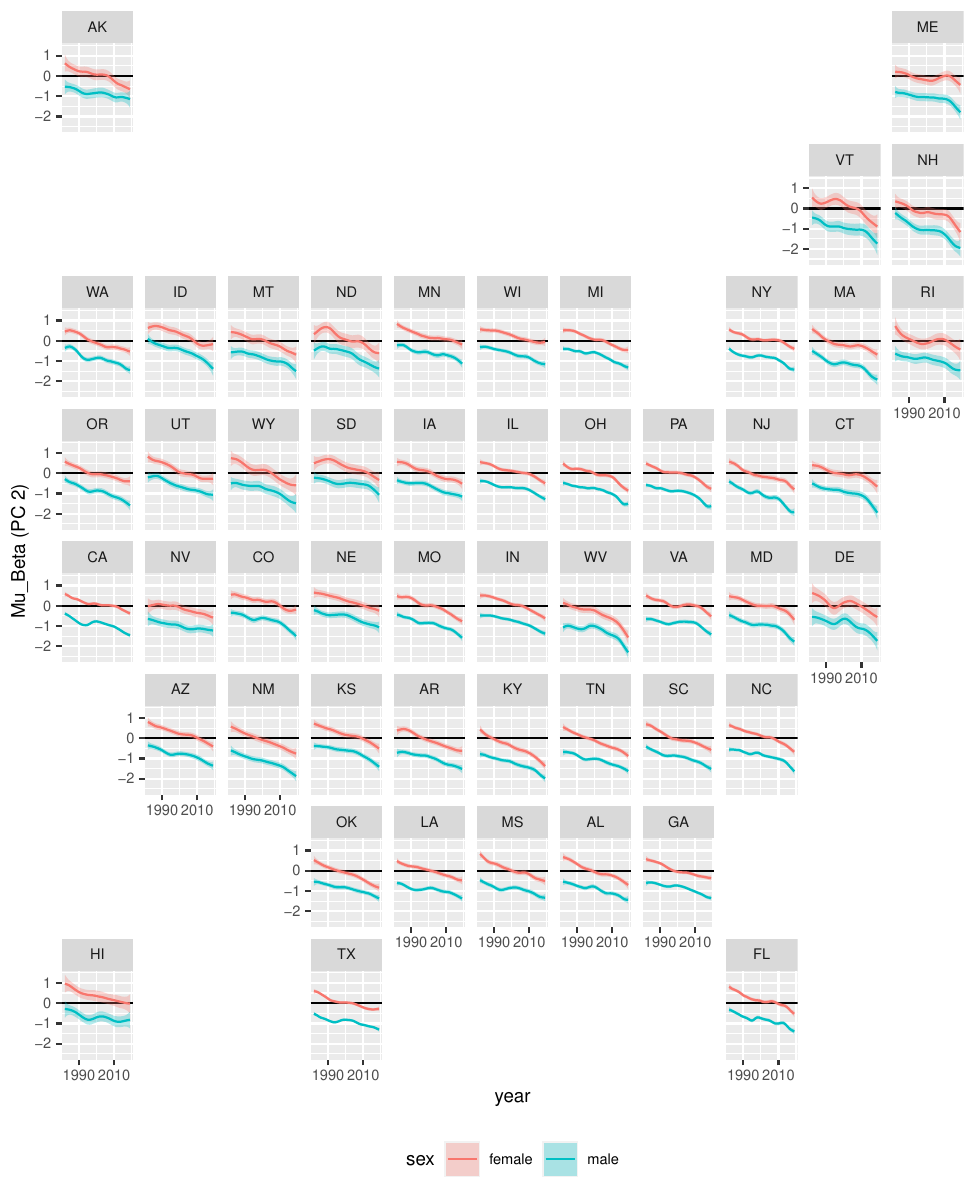} 

}

\caption{Posterior medians and 95\% credible intervals for the second principal component ($\mu_\beta$).}\label{fig:mubetamapPC2}
\end{figure}

In Figures \ref{fig:mubetamapPC1} and \ref{fig:mubetamapPC2}, we plot posterior medians and 95\% credible intervals of the estimated state-level coefficients for the first and second principal components for male and female populations in all states and years. States are organized roughly along their relative geographic location. The most notable finding is that in the plot for the first principal component, the gap in the baseline mortality coefficients between men and women is shrinking over time. This suggests a convergence in mortality across the sexes (Seligman, Greenberg, and Tuljapurkar 2016). Geographically, we see that baseline mortality patterns differ substantially by state, suggesting evidence for a divergence in mortality across US states due to stagnation or regression in some states and continuous improvement in others (Fenelon 2013). For example, states such as California and New Jersey experienced significant declines in baseline mortality in both sexes. Others such as Louisiana experienced limited declines and have stagnated in the 2010s. Some states such as Ohio saw a deterioration in baseline mortality in the 2010s. This trend reversal occurs across the eastern half of the US.

For the second principal component, declines are experienced across the US in both sexes with female values transitioning from positive to negative. As the second principal component contains an inverted accident hump shape, this suggests increasing young adult mortality relative to the baseline. Our findings suggest that this trend accelerated in the 2010s for males in New Jersey, West Virginia, Ohio, and more broadly across the Northeast and Midwest. This is consistent with the sharp increase in opioid overdose deaths in these regions of the US (Alexander, Kiang, and Barbieri 2018).

For the final two principal components (shown in Appendix \ref{app-res}) there are again clear geographic clusters of patterns over time. For the third principal component, which has relatively higher mortality in mid adult ages, trends are generally declining, but stagnating more in the South, and there is evidence of a trend reversal in Kentucky and West Virginia. Unlike the rest of the states, Alaska does not show any significant gap between the two sexes in the third principal component, suggesting that the deviation to baseline mortality associated with this component contributes similarly to male and female mortality. However, it is important to recognize that the population in Alaska is small, leading to large uncertainty in the extracted trends.

For the fourth principal component, a transition from positive to negative coefficient values is found across the US. The interpretation of this trend is similar to that of the second principal component. An inverted fourth principal component implies relatively higher early adult mortality, which is again consistent with the opioid crisis.

\begin{figure}

{\centering \includegraphics[width=.95\linewidth]{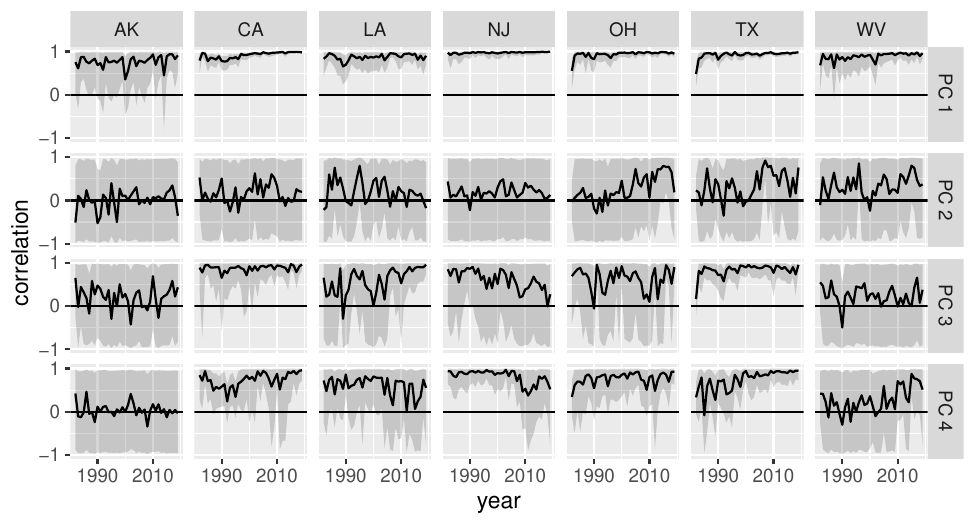} 

}

\caption{Time-series of posterior principal component correlations for seven states.}\label{fig:allbetacors}
\end{figure}

Figure \ref{fig:allbetacors} plots the posterior values of the between-sex principal component coefficient correlations extracted from \(L^{(\beta)}_{i,t} L_{i,t}^{(\beta)\top}\) for a subset of states over time. Plots for all states are given in Appendix \ref{app-res}. Each series includes the associated uncertainty intervals along with a horizontal line at zero. Across states, the first principal component has high correlation between the two sexes, reinforcing the notion that baseline mortality for both sexes move together. In contrast, there is a limited relationship in the second principal component. This is consistent with the idea that the accident hump is a predominantly male phenomenon, which would imply weak correlation between sexes. The third and fourth principal component correlations deviate by state. For Alaska, the relationships are weak at best, for California and Texas, there are strong or growing correlations, and New Jersey has declining correlations.

\begin{figure}

{\centering \includegraphics[width=.95\linewidth]{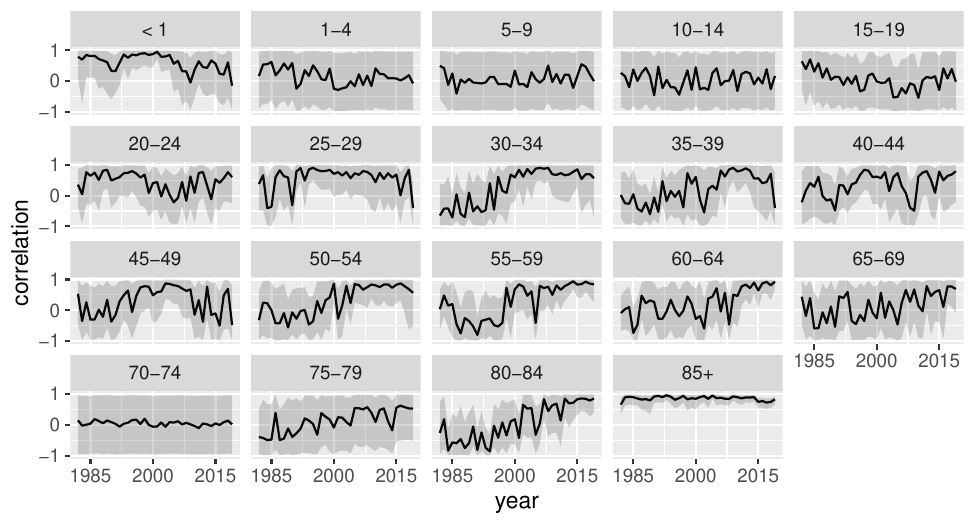} 

}

\caption{Posterior medians and 95\% credible intervals for $\gamma$ correlation over time by age group for California.}\label{fig:errs}
\end{figure}

In Figure \ref{fig:errs}, we plot the posterior medians and 95\% credible intervals for the between sexes correlations in the overdispersion terms captured by \(L_{a,t}^{(\gamma)}L_{a,t}^{(\gamma)\top}\) for California as an example. We find that patterns vary substantially by age group. For the 85+ category, consistent high correlation is found. However, in the youth age groups, there does not appear to be any meaningful correlation. The most interesting patterns are those in the mid adult age groups where transitions from limited correlation to strong positive correlation are found.

\hypertarget{validation}{%
\subsection{Validation}\label{validation}}

To formally evaluate the model, we perform out-of-sample validation exercises comparing the present model with the model proposed by Alexander, Zagheni, and Barbieri (2017). By comparing these two models, we can focus on the contributions of the between-subpopulation correlation matrices introduced. The validation exercises focus on five states (Alaska, California, Louisiana, New Jersey, and Texas) that capture the diversity in population size and number of counties across the US.

For each state, we leave out 20\% of the observed data for each county in the years 1982-2019 as an out-of-sample dataset, then execute the model on the remaining in-sample dataset. We then generate a distribution of deaths for all left-out observations using the posterior log-mortality rates for the corresponding county-age-sex-years. Finally, we calculate coverage at the 80\%, 90\%, and 95\% nominal levels for the death uncertainty intervals as well as mean squared errors (MSE) and mean absolute deviations (MAD) between the median death estimates and the observed deaths. In Table \ref{tab:val}, the results of this exercise by state are presented. Note that the model introduced in this paper is labeled as the ``joint'' model and the version in Alexander, Zagheni, and Barbieri (2017) with independent sex modelling is denoted ``independent''.

\begin{table}[!h]

\caption{\label{tab:val}Out-of-Sample Coverage and Errors}
\centering
\begin{tabular}[t]{llrrrrr}
\toprule
State & Model & Cov80 & Cov90 & Cov95 & MAD & MSE\\
\midrule
Alaska & independent & 0.884 & 0.943 & 0.973 & 1.931 & 14.447\\
\cellcolor[HTML]{E5E8E8}{Alaska} & \cellcolor[HTML]{E5E8E8}{joint} & \cellcolor[HTML]{E5E8E8}{0.887} & \cellcolor[HTML]{E5E8E8}{0.946} & \cellcolor[HTML]{E5E8E8}{0.974} & \cellcolor[HTML]{E5E8E8}{1.899} & \cellcolor[HTML]{E5E8E8}{12.614}\\
\midrule
California & independent & 0.843 & 0.920 & 0.961 & 9.747 & 2004.650\\
\cellcolor[HTML]{E5E8E8}{California} & \cellcolor[HTML]{E5E8E8}{joint} & \cellcolor[HTML]{E5E8E8}{0.846} & \cellcolor[HTML]{E5E8E8}{0.923} & \cellcolor[HTML]{E5E8E8}{0.963} & \cellcolor[HTML]{E5E8E8}{8.568} & \cellcolor[HTML]{E5E8E8}{1194.914}\\
\midrule
Louisiana & independent & 0.873 & 0.939 & 0.970 & 3.054 & 40.795\\
\cellcolor[HTML]{E5E8E8}{Louisiana} & \cellcolor[HTML]{E5E8E8}{joint} & \cellcolor[HTML]{E5E8E8}{0.874} & \cellcolor[HTML]{E5E8E8}{0.940} & \cellcolor[HTML]{E5E8E8}{0.972} & \cellcolor[HTML]{E5E8E8}{2.921} & \cellcolor[HTML]{E5E8E8}{30.257}\\
\midrule
New Jersey & independent & 0.843 & 0.925 & 0.963 & 8.388 & 298.082\\
\cellcolor[HTML]{E5E8E8}{New Jersey} & \cellcolor[HTML]{E5E8E8}{joint} & \cellcolor[HTML]{E5E8E8}{0.850} & \cellcolor[HTML]{E5E8E8}{0.930} & \cellcolor[HTML]{E5E8E8}{0.964} & \cellcolor[HTML]{E5E8E8}{7.838} & \cellcolor[HTML]{E5E8E8}{238.458}\\
\midrule
Texas & independent & 0.878 & 0.940 & 0.970 & 3.232 & 92.449\\
\cellcolor[HTML]{E5E8E8}{Texas} & \cellcolor[HTML]{E5E8E8}{joint} & \cellcolor[HTML]{E5E8E8}{0.879} & \cellcolor[HTML]{E5E8E8}{0.941} & \cellcolor[HTML]{E5E8E8}{0.970} & \cellcolor[HTML]{E5E8E8}{3.030} & \cellcolor[HTML]{E5E8E8}{61.618}\\
\bottomrule
\end{tabular}
\end{table}

The joint model consistently outperforms the independent model on the out-of-sample set across metrics. The level of outperformance is most pronounced in the larger states, notably California. The outperformance of the joint model on the out-of-sample set suggests that it may have uses beyond the US county context for estimating subnational mortality by subpopulation in jurisdictions without complete data.

\hypertarget{discussion}{%
\section{Discussion}\label{discussion}}

In this paper we extended principal component-based methods to jointly estimate subnational mortality across subpopulations. This approach leverages the inherent structural mortality patterns associated with the individual principal components and extracts correlations between groups to offer insight into the joint movements of mortality trends across groups. The model centers on a regression-based framework with four principal components that allow core patterns in age-specific mortality to be captured. The principal component coefficients are modelled hierarchically to allow for information exchange within states. County-specific effects on each component are assumed to be correlated across subgroups within a county. Our approach is validated using both a simulation study and with out-of-sample exercises using real subnational mortality data. We find that the proposed model is well calibrated and that its errors are smaller than those of the independent model in out-of-sample exercises.

We illustrate the model through estimating US county-level sex-specific mortality rates. An investigation into the parameters of the model highlights general trends in US mortality as well as state specific patterns. Notably, while movements in baseline mortality tend to manifest in both sexes simultaneously, specific features of the mortality curve such as elevated young adult mortality appear independent by sex. These results can offer direction to those working to reduce mortality pressures. For example, when correlations are high, aggregate policies may be effective. However, in jurisdictions where correlations are low or zero, more targeted policies may be necessary. State-level parameter estimates also highlighted clear geographic clustering of mortality patterns over time. Notably, there is evidence of stagnation in parts of the country, with increasing early adult mortality particularly in the eastern states, and a stagnating improvement in middle-age mortality in the southern states.

We identify two directions for future extensions to this work. The first is to study how the model performs in real-world circumstances similar to the simulation study where there are more than two subpopulations and the composition of the population is not approximately balanced across groups. For example, race/ethnicity-based mortality data in US counties. Many counties are dominated by one subgroup, and thus the frequency of low or zero death data is significantly higher than with aggregate or sex-specific data. The second direction is to apply the model in contexts where there are differences in mortality data collection or death registration coverage. In many countries, complete mortality data such as the US county data used here are not available and death registration coverage can vary substantially by sex, geographical area, or time (Peralta et al. 2019; Basu and Adair 2021). Typically, male coverage is higher than female coverage. By studying relationships between sexes in the years and regions with higher quality data, one may be able to support estimation of female mortality rates by using the male data and the modelled correlations.

\newpage

\hypertarget{references}{%
\section{References}\label{references}}

\hypertarget{refs}{}
\begin{CSLReferences}{1}{0}
\leavevmode\vadjust pre{\hypertarget{ref-AA}{}}%
Alexander, Monica, and Leontine Alkema. 2022. {``{A Bayesian Cohort Component Projection Model to Estimate Women of Reproductive Age at the Subnational Level in Data-Sparse Settings}.''} \emph{Demography} 59 (5): 1713--37. \url{https://doi.org/10.1215/00703370-10216406}.

\leavevmode\vadjust pre{\hypertarget{ref-alexander2018trends}{}}%
Alexander, Monica, Mathew V Kiang, and Magali Barbieri. 2018. {``Trends in Black and White Opioid Mortality in the United States, 1979--2015.''} \emph{Epidemiology (Cambridge, Mass.)} 29 (5): 707.

\leavevmode\vadjust pre{\hypertarget{ref-Alexander2017}{}}%
Alexander, Monica, Emilio Zagheni, and Magali Barbieri. 2017. {``A Flexible Bayesian Model for Estimating Subnational Mortality.''} \emph{Demography} 54. \url{https://doi.org/10.1007/s13524-017-0618-7}.

\leavevmode\vadjust pre{\hypertarget{ref-Arias2018}{}}%
Arias, Elizabeth, Loraine A. Escobedo, Jocelyn Kennedy, Chunxia Fu, and Jodi Cisewki. 2018. {``U.s. Small-Area Life Expectancy Estimates Project: Methodology and Results Summary.''} \emph{Vital and Health Statistics Series 2} 181. \url{https://www.cdc.gov/nchs/data/series/sr_02/sr02_181.pdf}.

\leavevmode\vadjust pre{\hypertarget{ref-barford2006life}{}}%
Barford, Anna, Danny Dorling, George Davey Smith, and Mary Shaw. 2006. {``Life Expectancy: Women Now on Top Everywhere.''} \emph{Bmj}. British Medical Journal Publishing Group.

\leavevmode\vadjust pre{\hypertarget{ref-india}{}}%
Basu, J. K., and T. Adair. 2021. {``{{H}ave inequalities in completeness of death registration between states in {I}ndia narrowed during two decades of civil registration system strengthening?}''} \emph{Int J Equity Health} 20 (1): 195.

\leavevmode\vadjust pre{\hypertarget{ref-HiramPNAS}{}}%
Beltrán-Sánchez, Hiram, Caleb E. Finch, and Eileen M. Crimmins. 2015. {``Twentieth Century Surge of Excess Adult Male Mortality.''} \emph{Proceedings of the National Academy of Sciences} 112 (29): 8993--98. \url{https://doi.org/10.1073/pnas.1421942112}.

\leavevmode\vadjust pre{\hypertarget{ref-Bhutta}{}}%
Bhutta, Zulfiqar A. 2016. {``{Mapping the geography of child mortality: a key step in addressing disparities}.''} \emph{The Lancet Global Health} 4 (12): E877--78. \url{https://doi.org/10.1016/S2214-109X(16)30264-9}.

\leavevmode\vadjust pre{\hypertarget{ref-pmid26902889}{}}%
Bijak, J., and J. Bryant. 2016. {``{{B}ayesian demography 250 years after {B}ayes}.''} \emph{Population Studies} 70 (1): 1--19. \url{https://doi.org/10.1080/00324728.2015.1122826}.

\leavevmode\vadjust pre{\hypertarget{ref-Clark}{}}%
Clark, Samuel J. 2019. {``A General Age-Specific Mortality Model with an Example Indexed by Child Mortality or Both Child and Adult Mortality.''} \emph{Demography} 56 (3). \url{https://doi.org/10.1007/s13524-019-00785-3}.

\leavevmode\vadjust pre{\hypertarget{ref-AnsleyBook}{}}%
Coale, Ansley J., Paul Demeny, and Barbara Vaughan, eds. 1983. Second Edition. Academic Press. \url{https://doi.org/10.1016/C2013-0-07295-7}.

\leavevmode\vadjust pre{\hypertarget{ref-crimmins2019aging}{}}%
Crimmins, Eileen M, and Yuan S Zhang. 2019. {``Aging Populations, Mortality, and Life Expectancy.''} \emph{Annual Review of Sociology} 45: 69--89.

\leavevmode\vadjust pre{\hypertarget{ref-deBeer}{}}%
de Beer, Joop. 2012. {``Smoothing and Projecting Age-Specific Probabilities of Death by TOPALS.''} \emph{Demographic Research} 27: 543--92. \url{http://www.jstor.org/stable/26349934}.

\leavevmode\vadjust pre{\hypertarget{ref-PTOPALS}{}}%
Dyrting, Sigurd. 2020. {``{Smoothing migration intensities with P-TOPALS}.''} \emph{Demographic Research} 43 (55): 1607--50. \url{https://doi.org/10.4054/DemRes.2020.43.55}.

\leavevmode\vadjust pre{\hypertarget{ref-Ezzati}{}}%
Ezzati, Ari B AND Kulkarni, Majid AND Friedman. 2008. {``The Reversal of Fortunes: Trends in County Mortality and Cross-County Mortality Disparities in the United States.''} \emph{PLOS Medicine} 5 (4): 1--12. \url{https://doi.org/10.1371/journal.pmed.0050066}.

\leavevmode\vadjust pre{\hypertarget{ref-GeoDiv}{}}%
Fenelon, Andrew. 2013. {``Geographic Divergence in Mortality in the United States.''} \emph{Population and Development Review} 39 (4): 611--34. \url{https://doi.org/10.1111/j.1728-4457.2013.00630.x}.

\leavevmode\vadjust pre{\hypertarget{ref-rhat}{}}%
Gelman, Andrew, and Donald B. Rubin. 1992. {``Inference from Iterative Simulation Using Multiple Sequences.''} \emph{Statistical Science} 7 (4): 457--72. \url{http://www.jstor.org/stable/2246093}.

\leavevmode\vadjust pre{\hypertarget{ref-Gjonca}{}}%
Gjonça, Arjan, Cecilia Tomassini, and James Vaupel. 1999. {``Male-Female Differences in Mortality in the Developed World.''} Rostock: Max Planck Institute for Demographic Research.

\leavevmode\vadjust pre{\hypertarget{ref-godwin2021space}{}}%
Godwin, Jessica, and Jon Wakefield. 2021. {``Space-Time Modeling of Child Mortality at the Admin-2 Level in a Low and Middle Income Countries Context.''} \emph{Statistics in Medicine} 40 (7): 1593--1638.

\leavevmode\vadjust pre{\hypertarget{ref-Schmertmann2}{}}%
Gonzaga, Marcos R., and Carl P. Schmertmann. 2016. {``Estimation of Mortality Rates by Age and Sex for Small Areas with TOPALS Regression: An Application for Brazil in 2010.''} \emph{Revista Brasileira De Estudos De Popula{ç}{ã}o} 33 (3): 629--52. \url{https://doi.org/10.20947/S0102-30982016c0009}.

\leavevmode\vadjust pre{\hypertarget{ref-gutin2021social}{}}%
Gutin, Iliya, and Robert A Hummer. 2021. {``Social Inequality and the Future of US Life Expectancy.''} \emph{Annual Review of Sociology} 47: 501--20.

\leavevmode\vadjust pre{\hypertarget{ref-harper2021declining}{}}%
Harper, Sam, Corinne A Riddell, and Nicholas B King. 2021. {``Declining Life Expectancy in the United States: Missing the Trees for the Forest.''} \emph{Annual Review of Public Health} 42: 381--403.

\leavevmode\vadjust pre{\hypertarget{ref-acchump}{}}%
Heligman, L., and J. H. Pollard. 1980. {``The Age Pattern of Mortality.''} \emph{Journal of the Institute of Actuaries} 107 (1): 49--80. \url{https://doi.org/10.1017/S0020268100040257}.

\leavevmode\vadjust pre{\hypertarget{ref-hendi2015trends}{}}%
Hendi, Arun S. 2015. {``Trends in US Life Expectancy Gradients: The Role of Changing Educational Composition.''} \emph{International Journal of Epidemiology} 44 (3): 946--55.

\leavevmode\vadjust pre{\hypertarget{ref-nuts}{}}%
Hoffman, Matthew D., and Andrew Gelman. 2014. {``The No-u-Turn Sampler: Adaptively Setting Path Lengths in Hamiltonian Monte Carlo.''} \emph{Journal of Machine Learning Research} 15 (47): 1593--623. \url{http://jmlr.org/papers/v15/hoffman14a.html}.

\leavevmode\vadjust pre{\hypertarget{ref-HMD}{}}%
Human Mortality Database. 2023. Max Planck Institute for Demographic Research (Germany), University of California, Berkeley (USA),; French Institute for Demographic Studies (France). \href{http://www.mortality.org\%20\%20(data\%20downloaded\%20on\%20August\%2025,\%202023)}{http://www.mortality.org (data downloaded on August 25, 2023)}.

\leavevmode\vadjust pre{\hypertarget{ref-Khana}{}}%
Khana, Lauren M. Rossen, Diba, and Margaret Warner. 2018. {``A Bayesian Spatial and Temporal Modeling Approach to Mapping Geographic Variation in Mortality Rates for Subnational Areas with r-INLA.''} \emph{Journal of Data Science} 16 (1): 147--82.

\leavevmode\vadjust pre{\hypertarget{ref-Kindig}{}}%
Kindig, David A., and Erika R. Cheng. 2013. {``Even as Mortality Fell in Most US Counties, Female Mortality Nonetheless Rose in 42.8 Percent of Counties from 1992 to 2006.''} \emph{Health Affairs} 32 (3): 451--58. \url{https://doi.org/10.1377/hlthaff.2011.0892}.

\leavevmode\vadjust pre{\hypertarget{ref-LeeCarter}{}}%
Lee, Ronald D., and Lawrence R. Carter. 1992. {``Modeling and Forecasting u. S. Mortality.''} \emph{Journal of the American Statistical Association} 87 (419): 659--71. \url{http://www.jstor.org/stable/2290201}.

\leavevmode\vadjust pre{\hypertarget{ref-LKJ}{}}%
Lewandowski, Daniel, Dorota Kurowicka, and Harry Joe. 2009. {``Generating Random Correlation Matrices Based on Vines and Extended Onion Method.''} \emph{Journal of Multivariate Analysis} 100 (9): 1989--2001. https://doi.org/\url{https://doi.org/10.1016/j.jmva.2009.04.008}.

\leavevmode\vadjust pre{\hypertarget{ref-masters2021changes}{}}%
Masters, Ryan, and Laudan Aron. 2021. {``Changes in US Life Expectancy in the Wake of COVID-19: Differences by Race/Ethnicity and Relative to Other High-Income Countries.''}

\leavevmode\vadjust pre{\hypertarget{ref-spacetime}{}}%
Mercer, Laina D., Jon Wakefield, Athena Pantazis, Angelina M. Lutambi, Honorati Masanja, and Samuel Clark. 2015. {``Space-Time Smoothing of Complex Survey Data: Small Area Estimation for Child Mortality.''} \emph{The Annals of Applied Statistics} 9 (December): 1889--1905. \url{https://doi.org/10.1214/15-AOAS872}.

\leavevmode\vadjust pre{\hypertarget{ref-montez2020us}{}}%
Montez, Jennifer Karas, Jason Beckfield, Julene Kemp Cooney, Jacob M Grumbach, Mark D Hayward, Huseyin Zeyd Koytak, Steven H Woolf, and Anna Zajacova. 2020. {``US State Policies, Politics, and Life Expectancy.''} \emph{The Milbank Quarterly} 98 (3): 668--99.

\leavevmode\vadjust pre{\hypertarget{ref-IHME}{}}%
Murray, Sandeep C AND Michaud, Christopher J. L AND Kulkarni. 2006. {``Eight Americas: Investigating Mortality Disparities Across Races, Counties, and Race-Counties in the United States.''} \emph{PLOS Medicine} 3 (9): 1--12. \url{https://doi.org/10.1371/journal.pmed.0030260}.

\leavevmode\vadjust pre{\hypertarget{ref-hmc}{}}%
Neal, Radford M. 2011. {``MCMC Using Hamiltonian Dynamics.''} In \emph{Handbook of Markov Chain Monte Carlo}. Chapman; Hall/CRC. \url{https://doi.org/10.1201/b10905}.

\leavevmode\vadjust pre{\hypertarget{ref-Noymer}{}}%
Noymer, Andrew, and Viola Van. 2014. {``{Divergence without decoupling: Male and female life expectancy usually co-move}.''} \emph{Demographic Research} 31 (51): 1503--24. \url{https://doi.org/10.4054/DemRes.2014.31.51}.

\leavevmode\vadjust pre{\hypertarget{ref-ecuador}{}}%
Peralta, A., J. Benach, C. Borrell, V. Espinel-Flores, L. Cash-Gibson, B. L. Queiroz, and M. Marí-Dell'Olmo. 2019. {``{{E}valuation of the mortality registry in {E}cuador (2001-2013) - social and geographical inequalities in completeness and quality}.''} \emph{Popul Health Metr} 17 (1): 3.

\leavevmode\vadjust pre{\hypertarget{ref-Raftery2014}{}}%
Raftery, Adrian E., Nevena Lalic, and Patrick Gerland. 2014. {``{Joint probabilistic projection of female and male life expectancy}.''} \emph{Demographic Research} 30 (27): 795--822. \url{https://doi.org/10.4054/DemRes.2014.30.27}.

\leavevmode\vadjust pre{\hypertarget{ref-DAE214715}{}}%
Rau, Roland, and Carl P. Schmertmann. 2020. {``{Lebenserwartung auf Kreisebene in Deutschland}.''} \emph{Dtsch Arztebl International} 117 (29-30): 493--99. \url{https://doi.org/10.3238/arztebl.2020.0493}.

\leavevmode\vadjust pre{\hypertarget{ref-Schmertmann1}{}}%
Schmertmann, Carl P., and Marcos R. Gonzaga. 2018. {``{Bayesian Estimation of Age-Specific Mortality and Life Expectancy for Small Areas With Defective Vital Records}.''} \emph{Demography} 55 (4): 1363--88. \url{https://doi.org/10.1007/s13524-018-0695-2}.

\leavevmode\vadjust pre{\hypertarget{ref-Sehgal}{}}%
Sehgal, Neil Jay, Dahai Yue, Elle Pope, Ren Hao Wang, and Dylan H. Roby. 2022. {``The Association Between COVID-19 Mortality and the County-Level Partisan Divide in the United States.''} \emph{Health Affairs} 41 (6): 853--63. \url{https://doi.org/10.1377/hlthaff.2022.00085}.

\leavevmode\vadjust pre{\hypertarget{ref-Seligman_34_38}{}}%
Seligman, Benjamin, Gabi Greenberg, and Shripad Tuljapurkar. 2016. {``{Convergence in male and female life expectancy: Direction, age pattern, and causes}.''} \emph{Demographic Research} 34 (38): 1063--74. \url{https://doi.org/10.4054/DemRes.2016.34.38}.

\leavevmode\vadjust pre{\hypertarget{ref-HS2021}{}}%
Ševčíková, Hana, and Adrian E. Raftery. 2021. {``Probabilistic Projection of Subnational Life Expectancy.''} \emph{Journal of Official Statistics} 37 (3): 591--610. \url{https://doi.org/doi:10.2478/jos-2021-0027}.

\leavevmode\vadjust pre{\hypertarget{ref-SONG2022102664}{}}%
Song, Insang, and Hui Luan. 2022. {``The Spatially and Temporally Varying Association Between Mental Illness and Substance Use Mortality and Unemployment: A Bayesian Analysis in the Contiguous United States, 2001--2014.''} \emph{Applied Geography} 140: 102664. https://doi.org/\url{https://doi.org/10.1016/j.apgeog.2022.102664}.

\leavevmode\vadjust pre{\hypertarget{ref-stan}{}}%
Stan Development Team. 2021. {``{RStan}: The {R} Interface to {Stan}.''} \url{https://mc-stan.org/}.

\leavevmode\vadjust pre{\hypertarget{ref-Stevenson}{}}%
Stevenson, John M., and David R. Olson. 1993. {``Methods for Analysing County-Level Mortality Rates.''} \emph{Statistics in Medicine} 12 (3-4): 393--401. https://doi.org/\url{https://doi.org/10.1002/sim.4780120320}.

\leavevmode\vadjust pre{\hypertarget{ref-USMDB}{}}%
United States Mortality Database. 2023. University of California, Berkeley (USA). \href{http://usa.mortality.org\%20(data\%20downloaded\%20on\%20August\%2025,\%202023)}{http://usa.mortality.org (data downloaded on August 25, 2023)}.

\leavevmode\vadjust pre{\hypertarget{ref-Wang2013}{}}%
Wang, Haidong, Austin E. Schumacher, Carly E. Levitz, Ali H. Mokdad, and Christopher JL Murray. 2013. {``Left Behind: Widening Disparities for Males and Females in US County Life Expectancy, 1985--2010.''} \emph{Population Health Metrics} 11 (8). \url{https://doi.org/10.1186/1478-7954-11-8}.

\leavevmode\vadjust pre{\hypertarget{ref-woolf2019life}{}}%
Woolf, Steven H, and Heidi Schoomaker. 2019. {``Life Expectancy and Mortality Rates in the United States, 1959-2017.''} \emph{Jama} 322 (20): 1996--2016.

\end{CSLReferences}

\newpage

\hypertarget{appendix-appendix}{%
\appendix}

\hypertarget{app-sim}{%
\section{Simulation study details}\label{app-sim}}

To illustrate the model's ability to estimate mortality rates and extract mortality patterns across subgroups, we simulate population and death data for populations composed of five subgroups of varying sizes. The simulation study data includes population and mortality data for 10 years, 25 subnational areas, and 5 population subgroups. For each year-area-age-subgroup, we construct the population data as follows:

\begin{enumerate}
\def\labelenumi{\arabic{enumi}.}
\tightlist
\item
  The total population in year 1 in each area is set as 100,000 times the area index, which ranges from 1-25.
\item
  The total population in each area in subsequent years is computed by increasing the total population in each area by 1\% annually.
\item
  Total populations in each year-area are allocated to the 19 age groups using a jittered version of the age distribution of Los Angeles County in California.
\item
  Each year-area-age population is then disaggregated into the 5 subgroups using the following proportions: 50\% in Group A, 20\% in Group B, and 10\% in Groups C, D, E.
\end{enumerate}

\begin{figure}

{\centering \includegraphics[width=.75\linewidth]{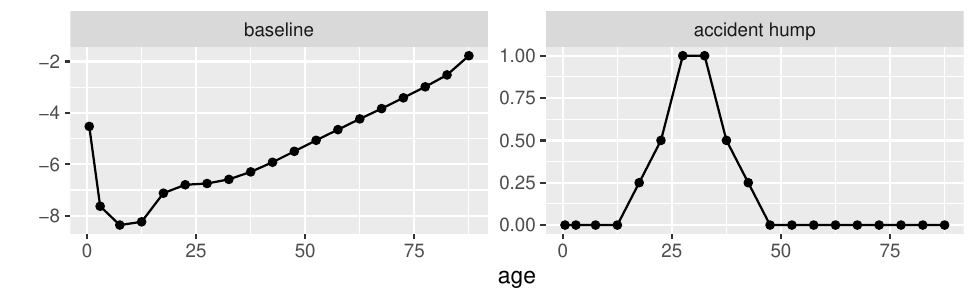} 

}

\caption{Simulation study mortality curve components}\label{fig:sim}
\end{figure}

To generate corresponding mortality data for the populations in each year-area-age-subgroup, we first construct true log-mortality curves for each year-area-subgroup. The mortality curves are constructed using a linear combination of two standard curves: the first representing a baseline mortality curve, and the second containing an accident hump. Plots of the standard curves are given in Figure \ref{fig:sim}.

The log-mortality curves are then constructed with the following steps:

\begin{enumerate}
\def\labelenumi{\arabic{enumi}.}
\tightlist
\item
  The baseline mortality curve is multiplied by a random coefficients sampled from a 5-dimensional multivariate normal distribution with mean \(\mathbf{1}_{5}\) and a year-dependent covariance matrix. The covariance matrix describes the dependence between the 5 subgroups. It is assigned a constant variance of \(0.1^2\) and one of the correlation structures plotted in Figure \ref{fig:simcor}a. In years 1-3, the independent correlation matrix is used. In years 4-6, the exchangeable correlation matrix (ie. constant off-diagonal values) is used. In years 7-10, the unstructured correlation matrix based on state-level observed race/ethnicity-based US mortality is used.
\item
  An accident hump is added to the mortality curve by adding the accident hump standard curve multiplied by random coefficients sampled from a 5-dimensional multivariate normal distribution with mean \(\mathbf{0}_{5}\), and covariance matrix with a constant variance of \(0.5^2\) and one of the correlation matrices plotted in Figure \ref{fig:simcor}a. The matrices used in each year mirror those used for the baseline mortality standard.
\end{enumerate}

Finally, observed deaths for each year-area-age-subgroup are generated from the log-mortality curves using a Poisson likelihood with rate equal to the corresponding population multiplied by the exponential of the corresponding log-mortality rate.

\begin{figure}

{\centering \includegraphics[width=.75\linewidth]{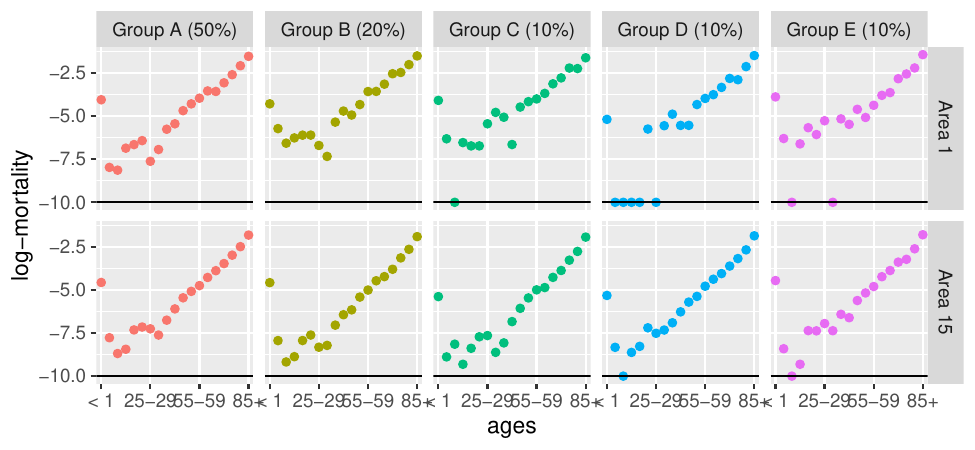} 

}

\caption{Examples of simulated log-mortality observations.}\label{fig:simdat}
\end{figure}

Examples of simulated data in two subnational areas are given in Figure \ref{fig:simdat}. Points correspond to log-mortality observations (ie. the log of observed deaths divided by population). Note that zero death observations are encoded as -10 on the log-scale.

\newpage

\hypertarget{app-res}{%
\section{Additional results}\label{app-res}}

\begin{figure}

{\centering \includegraphics[width=.75\linewidth]{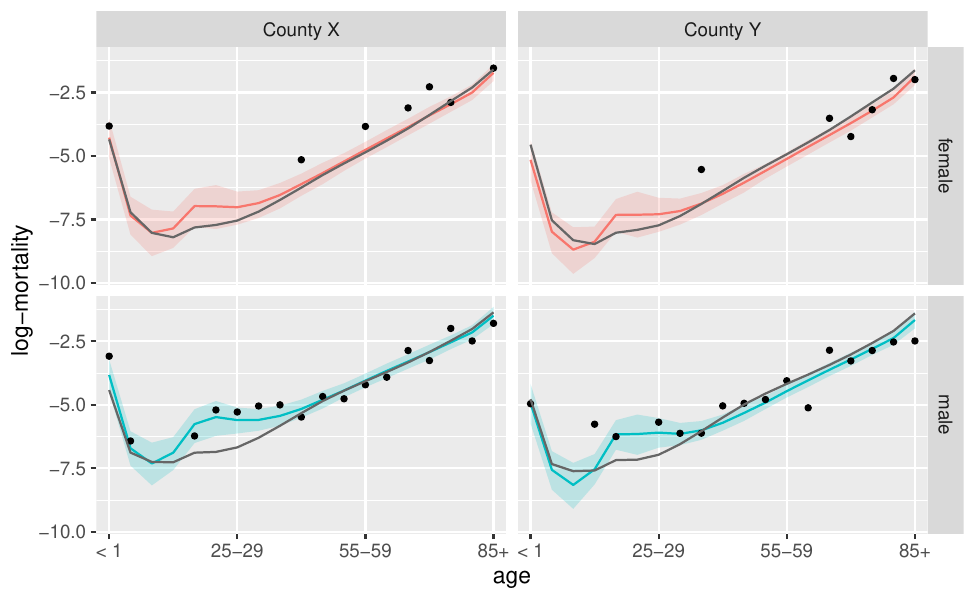} 

}

\caption{Examples of US counties with limited female but sufficient male data.}\label{fig:miss}
\end{figure}

Figure \ref{fig:miss} plots two county-years where the amount of non-zero death data varies by sex. The observed log-mortality rates for age groups with non-zero deaths are presented as points, the estimated state means for each sex are denoted by the grey lines, and the male and female county-level estimates and 95\% credible intervals are presented in blue and red respectively. In both county-years, the majority of age groups had zero female deaths in the year whereas the male data is nearly complete. As opposed to defaulting to the female state-level means, estimates for female mortality are also informed by the behaviour of male mortality relative to the male state-level means.

\begin{figure}

{\centering \includegraphics{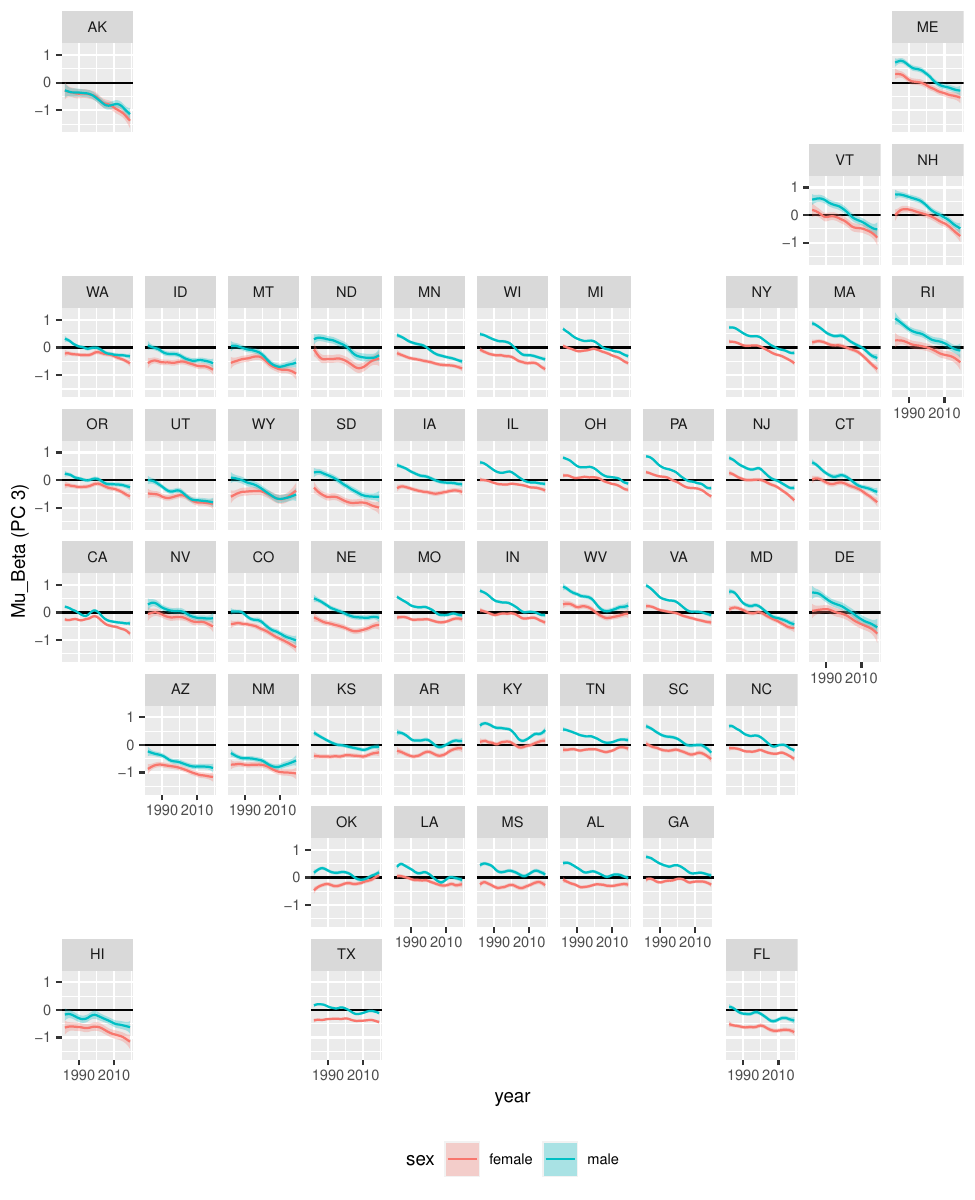} 

}

\caption{Posterior medians and 95\% credible intervals for the third principal component ($\mu_\beta$).}\label{fig:mubetamapPC3}
\end{figure}

\begin{figure}

{\centering \includegraphics{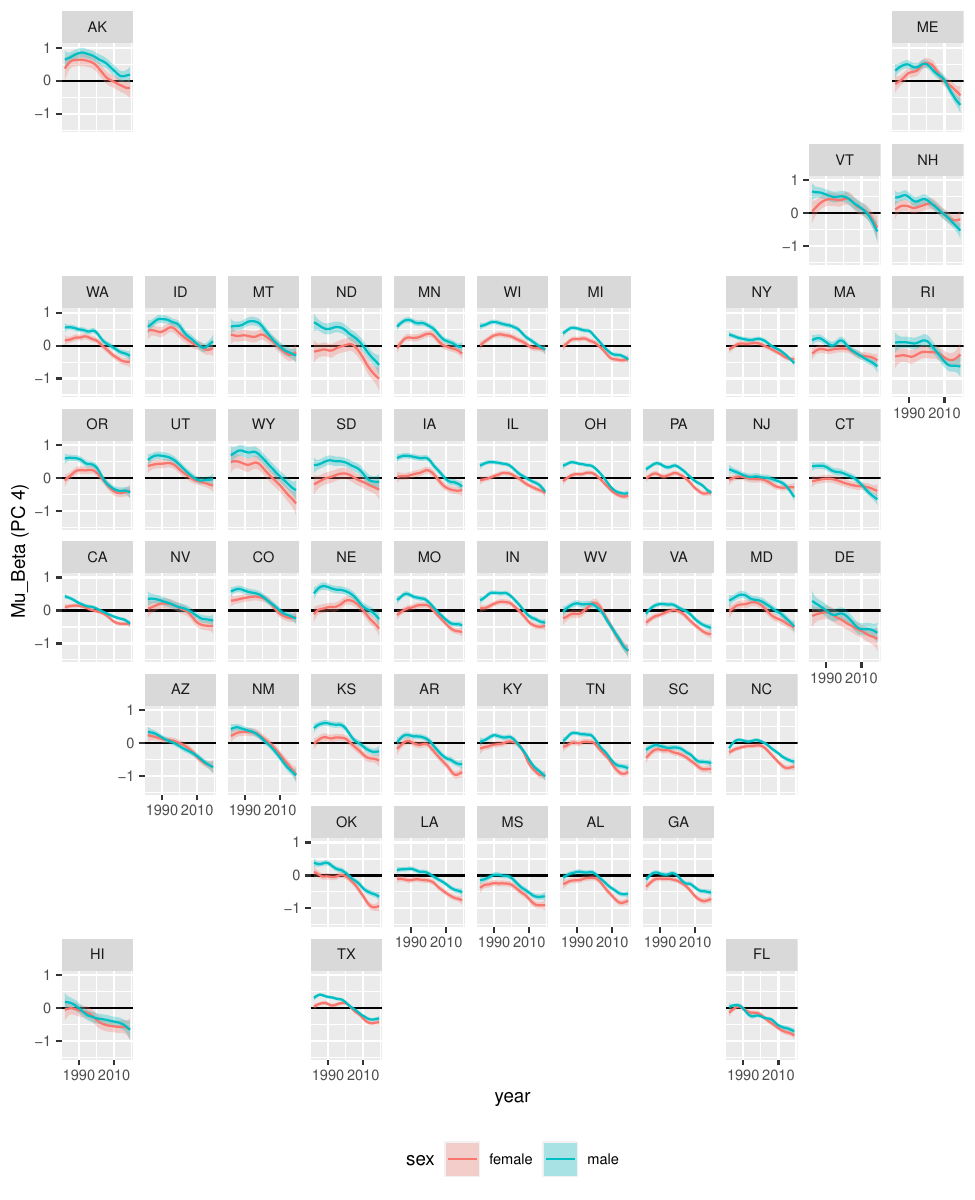} 

}

\caption{Posterior medians and 95\% credible intervals for the fourth principal component ($\mu_\beta$).}\label{fig:mubetamapPC4}
\end{figure}

Figures \ref{fig:mubetamapPC3} and \ref{fig:mubetamapPC4} plot maps of the posterior medians and 95\% credible intervals of the estimated state-level coefficients for the third and fourth principal components for male and female populations for all years.

\begin{figure}

{\centering \includegraphics[width=.95\linewidth]{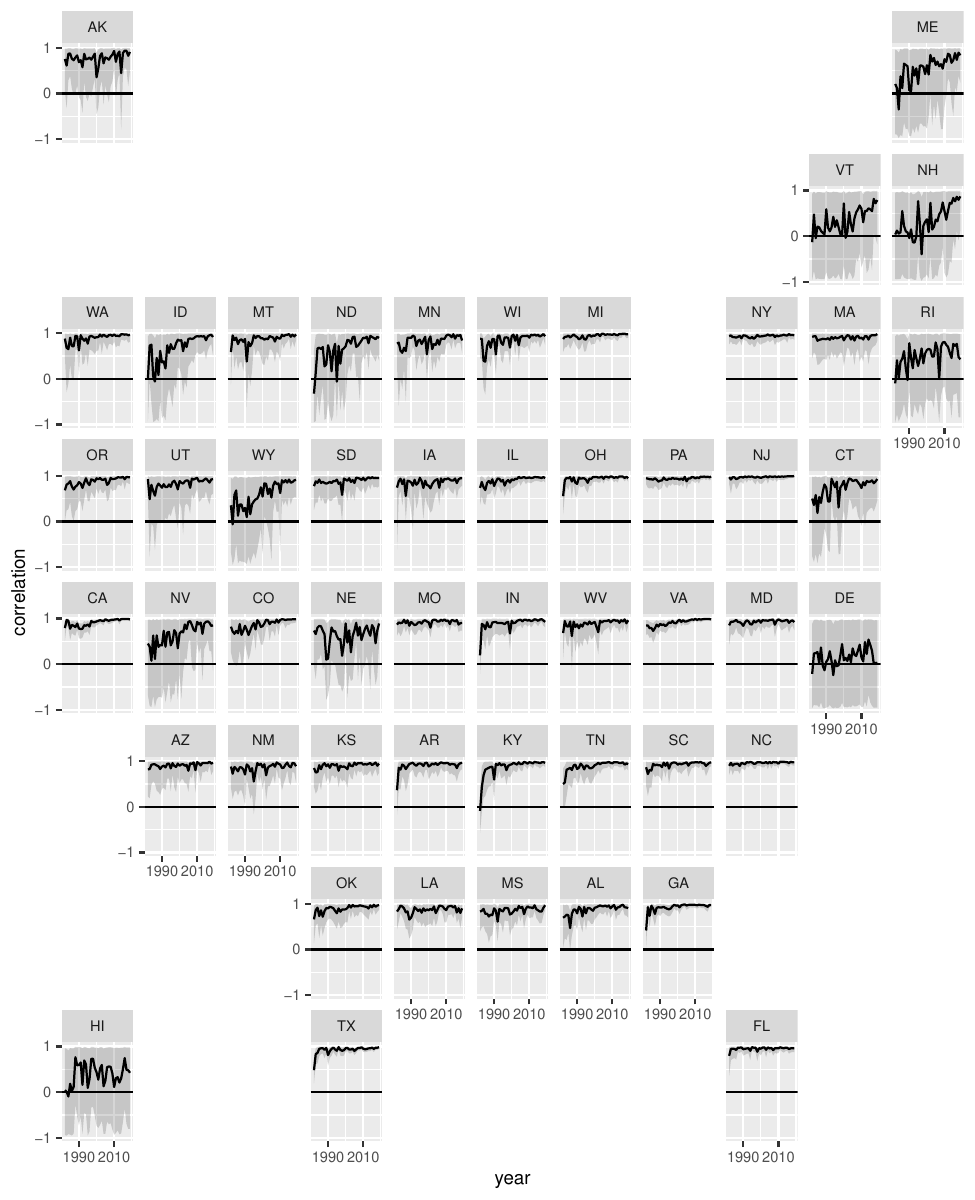} 

}

\caption{Time-series of posterior medians and 95\% credible intervals for the first principal component correlations.}\label{fig:betacors1}
\end{figure}

\begin{figure}

{\centering \includegraphics[width=.95\linewidth]{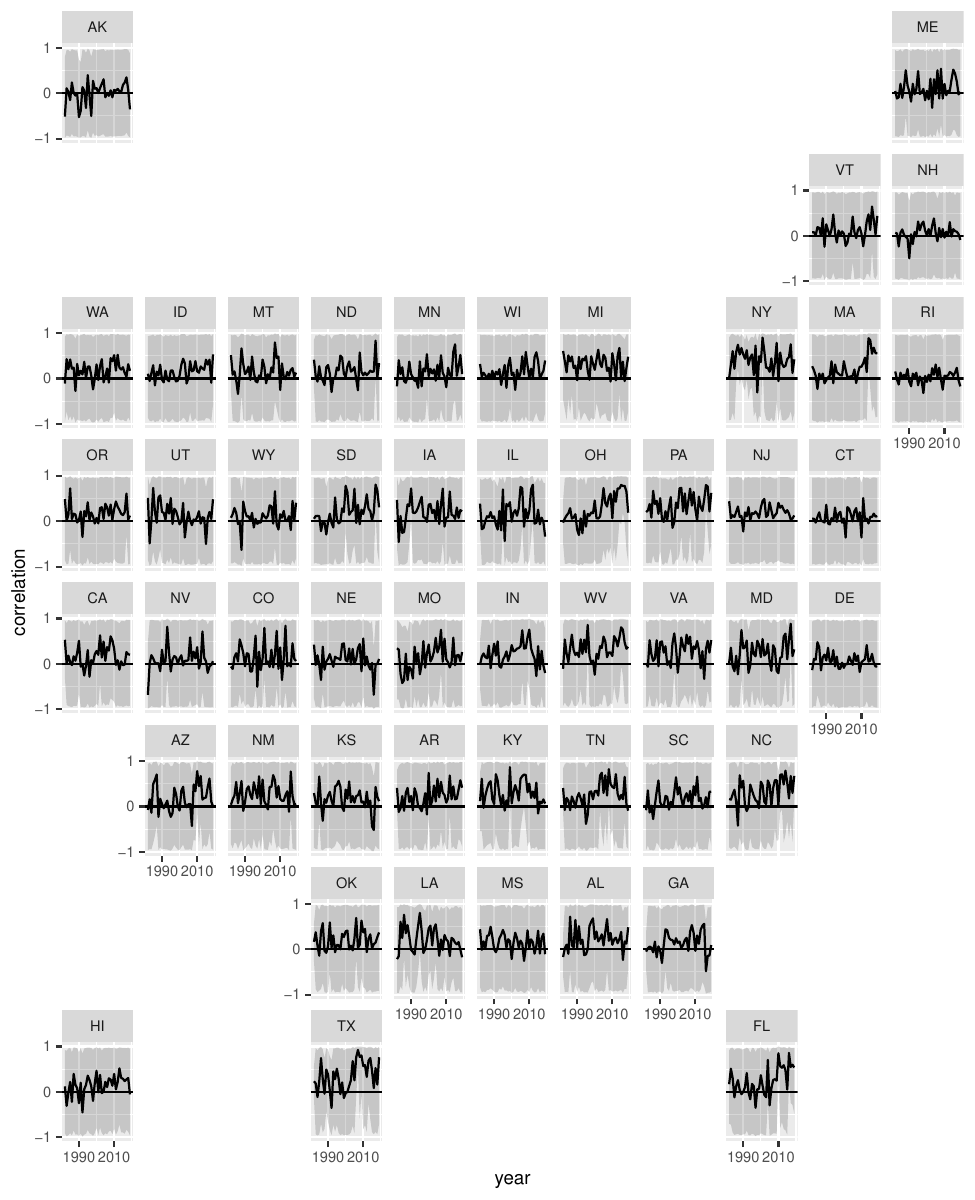} 

}

\caption{Time-series of posterior medians and 95\% credible intervals for the second principal component correlations.}\label{fig:betacors2}
\end{figure}

\begin{figure}

{\centering \includegraphics[width=.95\linewidth]{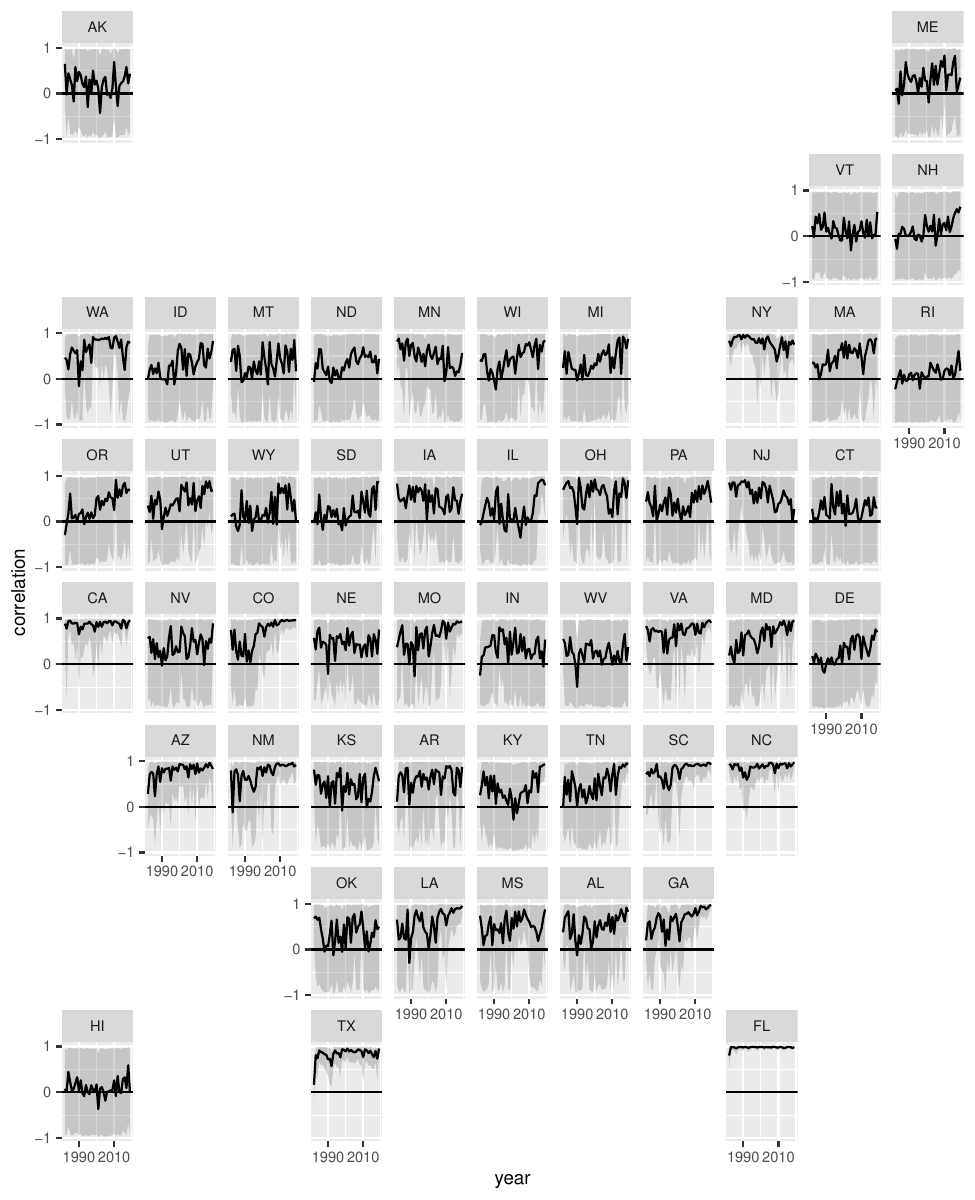} 

}

\caption{Time-series of posterior medians and 95\% credible intervals for the third principal component correlations.}\label{fig:betacors3}
\end{figure}

\begin{figure}

{\centering \includegraphics[width=.95\linewidth]{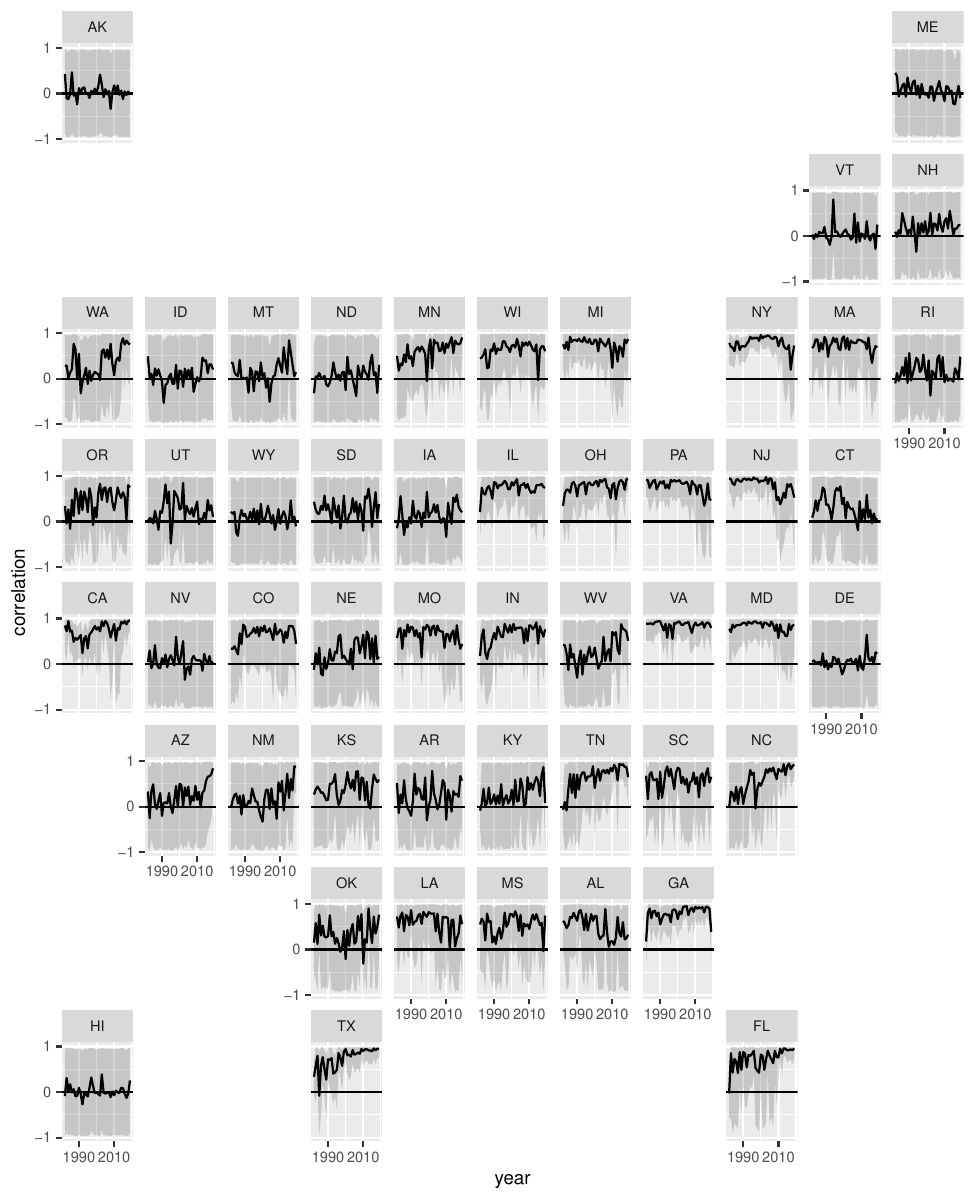} 

}

\caption{Time-series of posterior medians and 95\% credible intervals for the fourth principal component correlations.}\label{fig:betacors4}
\end{figure}

Figures \ref{fig:betacors1}, \ref{fig:betacors2}, \ref{fig:betacors3}, and \ref{fig:betacors4} plot the posterior medians and 95\% credible intervals for principal component coefficient correlations between sexes over time. As expected, strong correlation in the first principal component is found in most states whereas there is negligible correlation in the second principal component. Patterns in the third and fourth principal component coefficient correlations vary regionally with some evidence for stronger correlations in the fourth principal component coefficients in the Great Lakes region and surrounding states.

\end{document}